\documentclass{aa}  

\usepackage{graphicx}
\usepackage[flushleft]{threeparttable}
\usepackage{lmodern,babel,adjustbox,booktabs,multirow}
\usepackage{txfonts}
\usepackage{xcolor}
\begin{document}

   \title{Searching for cold gas traced by MgII quasar absorbers\\ 
   in massive X-ray-selected galaxy clusters}

    \titlerunning{Cold gas in galaxy clusters}

   \author{A. Y. Fresco
          \inst{1}
          C. P\'eroux\inst{2,3}\fnmsep
          A. Merloni\inst{1}
          J. Comparat\inst{1}
          R. Szakacs\inst{2}
          \and S. Weng\inst{2,4,5,6}
         }
   \institute{Max-Planck-Institut f\"ur Extraterrestrische Physik (MPE), Giessenbachstrasse 1, D--85748 Garching, Germany
         \and
             European Southern Observatory (ESO), 
             Karl-Schwarzschildstrasse 2, D-85748 Garching bei M{\"u}nchen, Germany
         \and
             Aix Marseille Universit\'e, CNRS, LAM (Laboratoire d'Astrophysique de Marseille) UMR 7326, 13388, Marseille, France 
         \and
             Sydney Institute for Astronomy, School of Physics A28, University of Sydney, NSW 2006, Australia 
         \and
             ARC Centre of Excellence for All Sky Astrophysics in 3 Dimensions (ASTRO 3D), Mount Stromlo Rd, Australian Capital Territory, 2611 Australia.
         \and
             ATNF, CSIRO Space and Astronomy, PO Box 76, Epping, NSW 1710, Australia                 
             }

   \date{May 10th, 2023}

 
  \abstract
{Almost 50\% of galaxies in the local Universe are in clusters or groups coexisting with both hot and cold gas components. In the present study, we observationally probed the cold-gas content of X-ray-selected massive galaxy clusters with spectroscopic redshift measured from the SDSS/SPIDERS survey. This paper focuses on the most massive structures: galaxy clusters with a mean mass of M$_{500c}$ =  2.7$\times 10^{14}$ M$_{\odot}$.} {We used a large number
of background quasar optical spectra from SDSS DR16 to probe the diffuse T$=$10$^4$K gas in their intracluster medium.} {We first analysed a sample of spectra with known MgII absorbers, and then blindly stacked about 16\,000 archival spectra at the redshifts of the foreground galaxy clusters.} {We tentatively ($3.7 \sigma$ significance) detect MgII in the clusters with an equivalent width EW(MgII $\lambda$2796) of 0.056$\pm$0.015  \AA, corresponding to a column density of log [N(MgII)/cm$^{-2}$]=12.12$\pm0.1$. We tested our methodology by generating 22\,000 mock SDSS spectra with MgII absorbers from TNG50 cosmological magnetohydrodynamical simulations, combining photo-ionisation modelling and ray tracing.  We also performed bootstrapping stacking at different cluster redshifts and stacked quasar spectra with no intervening clusters in the line of sight to measure the significance of our detection.} {These results are in line with the findings of recent, similar observational studies but challenge predictions from TNG simulations. Together, our findings indicate that large amounts of cold gas may be found in the most massive structures of the Universe.}

   \maketitle
%

\section{Introduction}

Galaxy clusters are an important laboratory for testing models of gravitational structure formation,  constraining the parameters of cosmological models, measuring the mean matter density of the Universe, and for investigating galaxy evolution as well as plasma physics \citep{Kravtsov12}. Despite the advances of decades of observations and simulations, the interactions between the gas phases that make up the baryon content of clusters remain largely unknown.

The baryon budget in clusters within R$_{500}$\footnote{We define R$_{500}$ as the radius where the density is 500 times the critical density of the Universe at a given redshift.}, although not well constrained, is made up of approximately 70\% hot intracluster medium (ICM), 13\% cold gas from stars and galaxies, and 17\% warm hidden baryons that are yet to be observed \citep{Ettori, Fukugita04, Gonzales07, Planelles13, Kravtsov05}.  The ICM in galaxy clusters has been thoroughly studied in X-rays with instruments such as ROSAT, XMM-Newton, Chandra \citep[see e.g.][and references therein]{Sarazin1986,Rosati2002,Boehringer2010}, and now eROSITA \citep{Merloni12, Predehl12, Liu2022}. In the meantime, optical and near-infrared observations from the Dark Energy Survey and the Hyper Supreme-Cam survey provide a different perspective on the growth and evolution of baryons in clusters \citep{Aihara18, DESI16}. Galaxy cluster outskirts are found to be multi-phase structures, where new infalling material flows inwards along filaments and cooler clumps become an important component \citep{Reiprich13}. The gas density distributions in this multi-phase structure depend on several physical properties, including temperature and dynamical state. Indeed, there exists a steep relation between gas mass and temperature that implies a decrease in the total gas mass content of cooler clusters relative to higher-mass systems \citep{Croston08}.

Alongside these developments, structure formation studies have progressively focused on the cycle of baryons. The physical processes by which gas is accreted onto galaxies, transformed into stars, and then expelled from galaxies into the circumgalactic medium (CGM) are of paramount importance for galaxy formation and evolution \citep{Tumlinson17}. The vast majority of the CGM studies so far have focused on field galaxies, while almost $\sim$50\% of galaxies in the local Universe are in clusters or groups \citep{Eke04}. In addition, simulations predict that a major fraction of the baryons at lower redshift are found in gas 10$^5$-10$^6$K, the so-called warm--hot intergalactic medium (WHIM; \citealp{CenOstriker99, Dave01}). Focusing on the Virgo cluster as a close-by laboratory, early results from \cite{Yoon} showed that the CGM surveys must consider the role of the environment. Clusters provide rather complex astrophysical and dynamical systems, where many different physical processes take place, such as galactic winds, active galactic nucleus (AGN) feedback, and gas stripping, leading to mixing and redistribution of metals in the ICM \citep{Simionescu09, K&M}. Nevertheless, as mentioned above, the baryon budget in clusters is not well constrained, which is mainly because it is unclear whether or not all baryonic constituents have been identified and quantified.

The ICM is an essential component for the assessment of the cosmic baryon and metal budgets, given that the ICM contributes even more to the overall baryon budget than stars \citep[e.g.][]{Gonzaleznov13,PerouxHowk20}. Within these large structures, the ICM material is composed and mixed with the CGM material of many group members. Galaxies are infalling and orbiting within halos, while their properties are likely affected by encounters with the warm and hot gas sitting in the gravitational well  \citep[e.g.][]{Popesso15}. The large amounts of hot gas trapped within their deep potential wells make clusters of galaxies shine in the X-ray waveband \citep[e.g.][]{JonesForman99}. However, determining the abundance of elements other than iron probed by X-ray observations is currently challenging and uncertain \citep{Mernier18, Frebel18}. 

From a theoretical standpoint, multi-dimensional numerical plasma simulations in the ICM of clusters, groups, and massive galaxies show how cold-gas filaments can condense out of a hot halo. These filaments fall through the hot gas where cool blobs can survive the passage through the hot medium due to Kevin-Helmholtz instability \citep{Sharma12, Mccourt12}. Modern cosmological hydrodynamic simulations show that cool--warm gas from the CGM of galaxies is stripped from the galactic potentials through ram-pressure forces of the hot ICM \citep{Ayromlou19, Yun19}. On the other hand, cosmic web filaments interact with galaxy clusters permeating gas streams of relatively metal-poor cool--warm gas. These and other mechanisms create an inhomogeneous and turbulent multi-phase ICM with a range of physical and chemical properties \citep{Bahe13,Simionescu19,Kravtsov12, Kunz22}. Models of interactions between the supermassive black holes and the large-scale atmospheres from these massive structures suggest that the cooling of the CGM is rapid and inhomogeneous \citep{Donahue22}. The conditions of the ambient gas in these haloes quickly develop a multi-phase structure, where the colder and denser gas sinks into the central galaxy and is accreted by the supermassive black hole \citep{Tremblay16}. This cold gas is then again pushed outwards through AGN outflows, but can rain back down after the hydrodynamic phenomena that lifted it subsides \citep{Mccourt12, Sharma12}. This cycle has been observed and was interpreted as being driven by AGN feedback \citep{Revaz08}.

The hotter component (T $\sim$ 10$^{6-7}$K) of this multi-phase gas emits at X-ray wavelengths. As a reference, the hot phase of the Milky Way's CGM has been observed both in emission and absorption, and has been characterised in great detail and found to be made up of three distinct components: the hot, warm, and warm--hot components \citep{Das21,Mathur21,Locatelli2023}. Beyond the Milky Way observations, the gas is diffuse and challenging to detect in emission due to low surface brightness \citep{Cantalupo19}; it is best probed using quasar absorption line spectroscopy, which has proven to be a powerful probe of these environments \citep{PerouxHowk20}. Absorption lines detected against bright background quasars offer the most compelling way to study the distribution, chemical properties, and kinematics of CGM gas \citep{Hamanowicz20, Roland21}. In these quasar absorbers, the minimum gas density that can be detected is set by the brightness of the background source and thus the detection efficiency is independent of redshift \citep{Tripp98}. While individual absorption measurements are limited to a pencil-beam along the line of sight, a sample of sightlines allows us to statistically measure the mean properties of galaxy clusters. These techniques have been extensively used to investigate the gaseous halos of isolated galaxies \citep{Roland21}. However, without a quantitative description of the most massive structures in the low-redshift Universe, a full census of baryons and metals in the Universe cannot be obtained.

Among quasar absorbers, the low-ionisation MgII doublet is known to trace cold 10$^4$K gas. Because of its distinct doublet feature, MgII has been used extensively in a large number of spectroscopic surveys. In the last two decades, MgII absorption system surveys have been used to study the physical properties of large samples of galaxies over a wide range of luminosities and morphologies \citep{Lanzetta90, Nestor05, Narayan07, Lopez, Seyffert13, Anand21}. Absorption by MgII in cosmological
galaxy formation simulations has been analysed \citep{Nelson20, Augustin21}, and is shown to closely relate to star forming regions, galactic outflow, and galactic discs \citep{Bowen}. These are known to be associated with galaxies and their CGM \citep{bouche, Zhu15}, and individual absorption lines allow us to characterise the spatial distribution and physical properties of the cold-gas clouds located in the vicinity of galaxies \citep{Lan&Mo18, Zhu15}.

Recently, large spectroscopic surveys have focused on the incidence of MgII in and around galaxy clusters, with different methods and samples, showing a wide range of detections \citep{Anand22, Mishra}. In particular, \citet{Mishra23} probed cold neutral gas in the outskirts of low-redshift galaxy clusters. The present study extends previous works to higher cluster masses. Indeed, we build on these works to analyse the cold gas traced by MgII absorbers in a large sample of massive X-ray-selected clusters. Aiming at this higher mass range provides fresh clues as to the evolution of the physical properties of galaxy clusters, including gas mass fractions as a function of galaxy cluster masses.

The present paper is organised as follows: Section 2 presents observational data used in this study. Section 3 details the analysis performed, while Section 4 focuses on the results. In section 5, we discuss our findings in the context of other works and state-of-the-art simulations. We summarise our findings and present our conclusions in Section 6. Here, we adopt an H$_0$ = 67.74 km s$^{-1}$ Mpc$^{-1}$, $\Omega_M$ = 0.3089, and $\Omega_{\Lambda}$ = 0.6911 cosmology. The equivalent width (EW) always refers to the MgII 2796\AA\ line unless stated otherwise.

\section{Observational data}

\begin{figure}
\centering
\includegraphics[width=1\columnwidth]{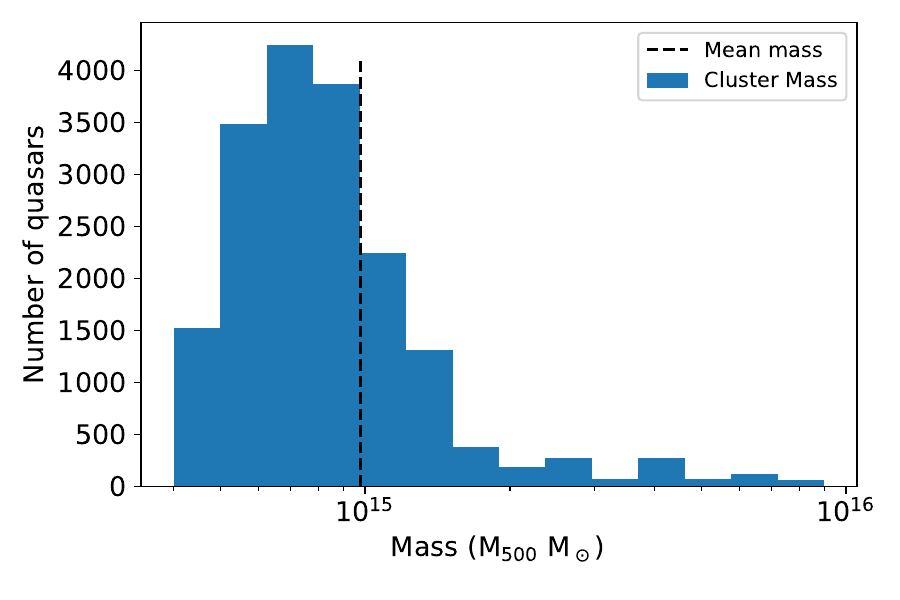}
\caption{Distribution of the number of background quasar spectra of the full sample used in the analysis as a function of the foreground cluster mass within R$_{500}$. The dotted line represents the mean mass of the clusters.}
\label{fig:mass}
\end{figure}

\subsection{The foreground cluster sample}

We start our analysis from the largest spectroscopically confirmed sample of massive clusters compiled from the SPectroscopic IDentification of ERosita Sources (SPIDERS) program of the Sloan Digital Sky Survey (SDSS; \citealp{Clerc2016,Clerc2020,Kirkpatrick21, IderChitman20}). We drew the cluster candidates for SPIDERS from a subset of CODEX \citep{Finoguenov20}, an X-ray-selected catalogue of clusters from the Rosat All Sky Survey (RASS), specifically in the 5\,350 square degree BOSS imaging footprint; 2\,740 CODEX clusters are included in the SPIDERS sample \citep{Kirkpatrick21}. These are complemented with the X-CLASS cluster sample (based on serendipitous XMM-Newton observations) with 124 uniquely identified and confirmed clusters \citep{Clerc2014}. Visual inspections of individual spectra are carried out by trained astronomers to verify the existence of a cluster. After a first run with the \textsc{redMaPPer} \citep{redmapper} algorithm for automatic membership assignment, a minimum of three visual inspecting members are required to converge on a final redshift determination, the final condition being that these members lie close to each other in a velocity--distance diagram. Due to the uncertainty associated with the position of the detections, several measures are taken to ensure the closest possible estimation of redshift.
The likelihood that a candidate will be validated is highly dependent on redshift, cluster richness, and the number of spectra available. For example, in the spectroscopic redshift determination, there are 622 instances where the mean spread in the measurements is $\delta_z$ = 0.00049 (147 km s$^{-1}$), while the maximum is $\delta_z$ = 0.0055 (1 650 km s$^{-1}$). We conservatively assume that all the objects with velocity offsets from the cluster centre greater than 5000 km s$^{-1}$ are rejected as members of the cluster. 

\begin{figure}
\centering
\includegraphics[width=0.9\columnwidth]{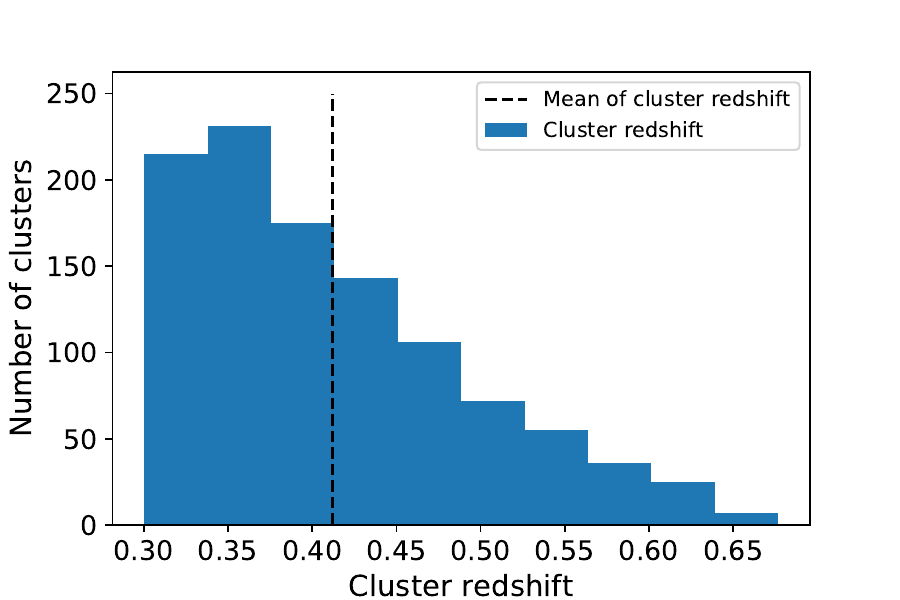}
\caption{Distribution of cluster redshifts from the parent SPIDERS sample \citep{Clerc2020} used in this study. The clusters constitute a subsample of 1066 objects from the parent sample of 2740 for which MgII absorption lines would fall within the observed SDSS spectral range. The dotted line represents the mean redshift of the clusters at z=0.41.}
\label{fig:clusz}
\end{figure}

 Figure~\ref{fig:mass} shows the distribution of foreground cluster masses. Scaling from their X-ray luminosities, the average mass of these systems amounts to M$_{500c}$ =  2.7$\times 10^{14}$ M$_{\odot}$ (converted from M$_{200c}$ =  3.9$\times 10^{14}$ M$_{\odot}$). Here, we mainly express the cluster mass in units of M$_{200c}$, that is, the mass within a radius R$_{200c}$. This corresponds to the radius enclosing an average density that is 200 times the critical density of the Universe at the cluster's redshift and ranges within  0.048 $<$ R$_{200c}$ $<$ 0.160 deg or within 1Mpc$<$ R$_{200c}<$ 3Mpc depending on redshift. This radius is expressed as:

\begin{equation}
    \mathrm{R_{200}}= \bigg(\frac{3\mathrm{M_{200c}}}{4\pi 200 \rho_{cr}} \bigg(\frac{\mathrm{H_{z}}}{100} \bigg)^{-2} \bigg)^{1/3}
    \label{eqn:ClusterMass}
.\end{equation}

In order to compare these results to the latest related published research, we use the public open-source package \textsc{COLOSSUS} to convert between M$_{200c}$ and M$_{500c}$ \citep{Diemer18}, and continue to do so throughout the paper for consistency. 

The observed galaxy cluster distribution sample covers a redshift span of 0.03 $<$ z $<$ 0.677 
(a subset of which ---systems with z$>$0.3--- is shown in Figure~\ref{fig:clusz}), with a typical statistical uncertainty on each redshift of $\Delta_{z}/(1+z)= 6 \times 10^{-4}$ \citep{Kirkpatrick21}. The number of spectroscopic members per system ranges between 3 and 75, with a mean of 12 members. 

\subsection{The background quasar sample}

We selected spectra of background quasars from SDSS. The spectra were obtained with the 2.5m Sloan Telescope as part of the BOSS and eBOSS surveys \citep{Smee13}, covering a wavelength range of 3600 \AA\ to 10\,400 \AA\ at a spectral resolution of $\lambda$/$\Delta \lambda \sim$2000, or with the SDSS-I instrument, covering 3800 \AA\ to 9100 \AA, with the same spectral resolution. The Data Release 16 of SDSS comprises a complete selection of spectroscopically confirmed quasars \citep{SDSSLyke, DR16}, and includes 750\,414 confirmed quasars with 0.8 $<$ z $<$ 2.2 (Table \ref{tab:numberspectra}).

\subsection{Pairing background quasars with foreground clusters}

\begin{figure}
\centering
\includegraphics[width=1\columnwidth]{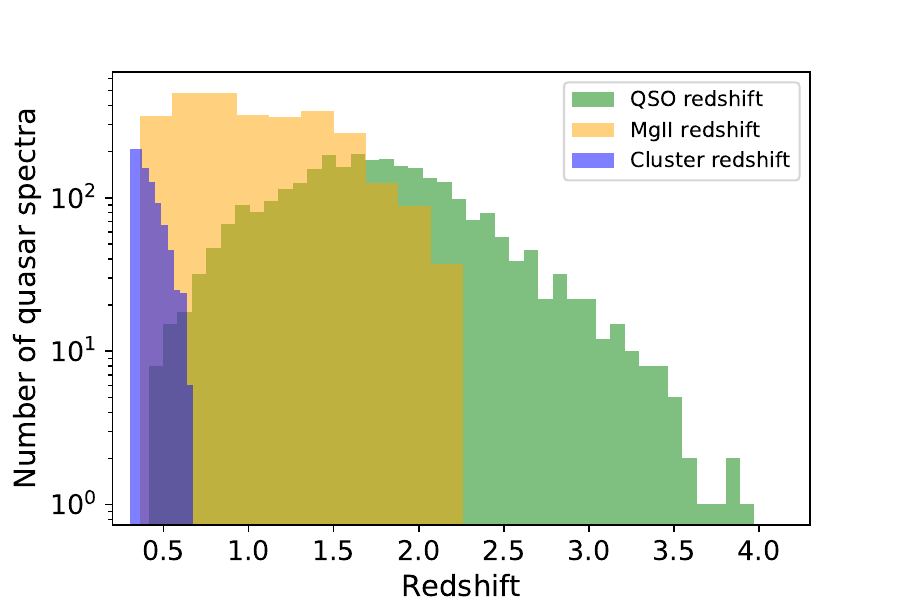}
\caption{Histogram of the number of quasars as a function of their redshift. Here, the green colour represents the quasar redshift (16,224 quasars), the orange colour represents the 2,881 MgII known absorbers redshift from \protect\cite{Anand21} found in these quasars, and the blue as the redshift of the foreground clusters (1,066 clusters).}
\label{fig:Zhist}
\end{figure}

\begin{figure}
\centering
\includegraphics[width=1\columnwidth]{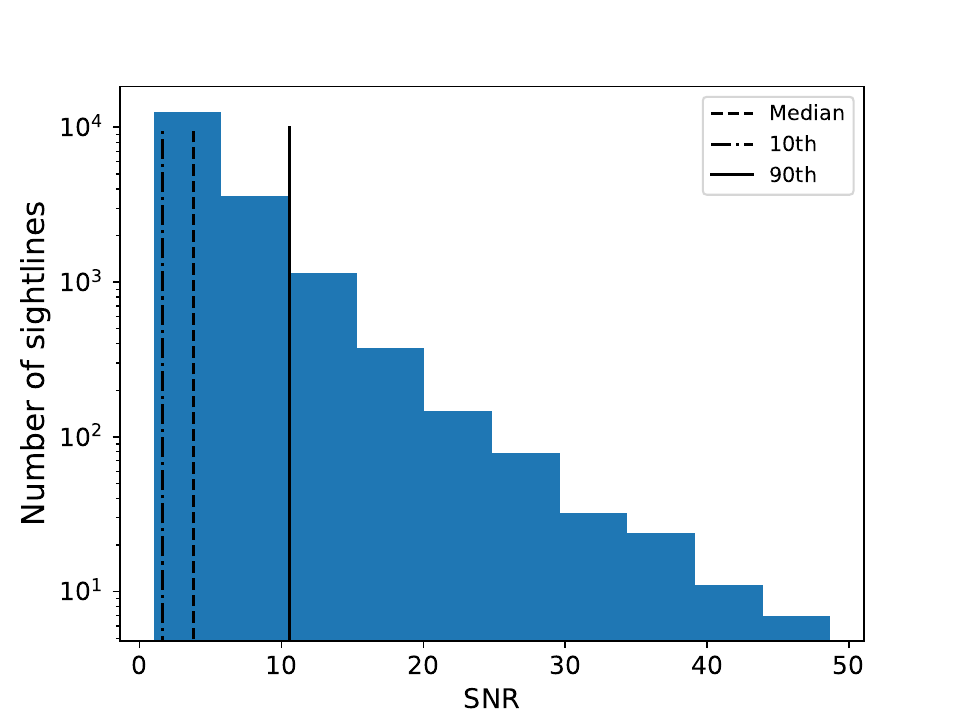}
\caption{Distribution of the S/N of the spectrum of 16,224 background quasars (see text) from the SDSS DR16 sample with foreground clusters. The median S/N of spectra is S/N$\sim$3.79 as indicated by the vertical dotted line. The 10$^{th}$ percentile is a dashed-dotted line, and the 90$^{th}$ percentile is the solid line.}
\label{fig:snr}
\end{figure}

A key component of the analysis is to determine which quasar sightlines pass close to a foreground galaxy cluster in projection on the plane of the sky. We determined the number of quasar--cluster pairs by cross-matching the foreground cluster sample with the background quasar catalogue. We made use of the cluster optical centre from the DR16 catalogue. We selected all SDSS spectra from the parent sample with a distance in right ascension and declination within three times R$_{200}$  (apparent R$_{200c}$, $\sim 3$ Mpc) of the centre of a cluster, and a redshift of z$_{\rm quasar}$ $>$ z$_{\rm cluster}$ + 0.01 for a total of 434,736 spectra. Among these cluster--quasar pairs, we select the ones for which the foreground cluster will have a redshift such that MgII is covered by the SDSS spectra (z$>$0.3), and a quasar redshift of  z $<$ 5. This leads to a total of 18\,694 cluster--quasar pairs. Table \ref{tab:numberspectra} presents the numbers of objects resulting from the various selection cuts: (i) the quasar redshift must be higher than the redshift of the intervening cluster; (ii) the corresponding MgII absorption line should fall within the SDSS wavelength coverage and the quasar redshift must be z $<$ 5; and (iii) the final spectrum must be free of wavelength gaps and other errors. Indeed, some of the SDSS spectra have gaps in their wavelength coverage due to bad detector columns or high noise peaks related to cosmic rays for example. These objects were removed from the sample. Figure~\ref{fig:Zhist} shows the distribution of the redshift of the High-quality SDSS spectra, the redshift of the known MgII absorbers in these spectra, and the redshift of the foreground clusters. We note that the cluster centre is not always well constrained because of the uncertainty in the redshift of some of the members.

\begin{table}[!ht]
\begin{threeparttable}
\begin{adjustbox}{width=.48\textwidth}
    \begin{tabular}{lc}
    \hline \hline 
    Sample & Number of objects\\
    \hline
       Parent sample  & 750,414 \\
       z$_{\mathrm{quasar}}$ > z$_{\mathrm{cluster}}$ and up to 3$\times$R$_{200c}$  & 434,736  \\
       MgII range and z$_{\mathrm{quasar}}$ $< 5$ & 18,694\\
       High-quality SDSS spectra  & 16,224 \\
    \hline \hline 
    \end{tabular}
    \end{adjustbox}
    \caption{Number of spectra resulting from the various selection cuts.}
    \footnotesize{Table \ref{tab:numberspectra} cuts: (i) the quasar redshift must be higher than the redshift of the intervening cluster; (ii) the corresponding MgII absorption line should fall within the SDSS wavelength coverage and the quasar redshift must be z $<$ 5; and (iii) the final spectrum must be free of wavelength gaps and other errors.}
    \label{tab:numberspectra}
\end{threeparttable}
\end{table}



The signal-to-noise ratio (S/N) is computed following \cite{Mas-Ribas17} for each spectrum as S/N= mean(flux)$/$spread, where the spread is the standard deviation defined as the dispersion of the flux relative to the mean, and the mean calculated over the flux of the spectra. The distribution of the S/N of the spectra is presented in Figure~\ref{fig:snr}. 

For each of the pairs, we calculate the angular separation between the foreground cluster and the background quasar projected on the sky plane. The angular separation is normalised by the R$_{200}$ radius of the cluster. We stress that while we select in units of R$_{200}$, all the figures presented in the present paper are in units of R$_{500}$ in order to ease comparison with results from the literature.

\subsection{ Sample of known MgII quasar absorbers}

Our analysis also makes use of the MgII quasar absorption catalogue of \cite{Anand21}. The authors developed an automated pipeline to detect intervening metal absorption line systems with a matched kernel convolution technique and adaptive S/N criteria. By processing one million quasars from the SDSS DR16, these authors compiled a sample of about 160\,000 MgII absorbers in the redshift range 0.3 $<$ z$_{\rm abs}$ $<$ 2.3. After cross-matching our sample of cluster--quasar pairs with this absorption catalogue, we get a match for 4\,150 spectra.

\begin{figure}
\centering
\includegraphics[width=1\columnwidth]{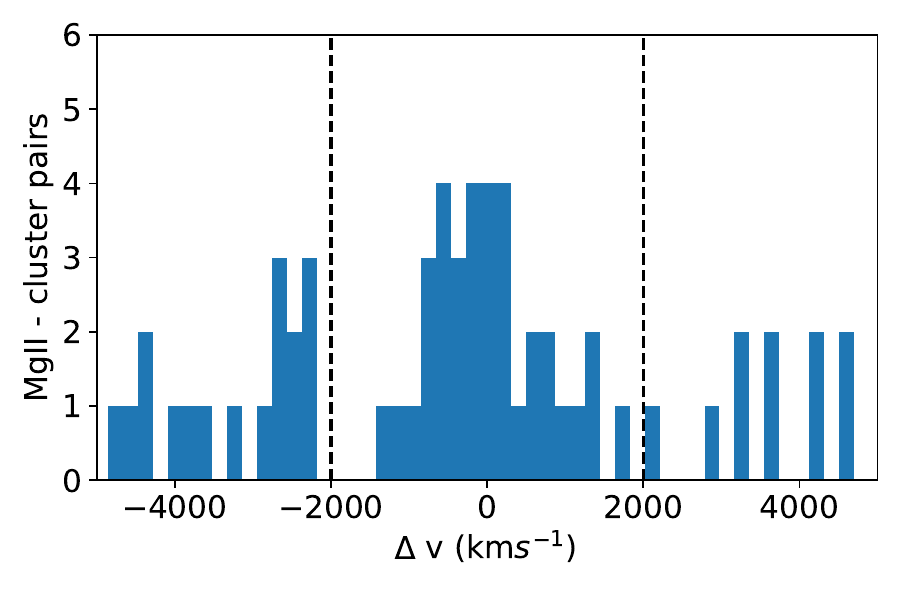}
\caption{Velocity difference, $\Delta$v, between the known MgII absorber from \protect\cite{Anand21} within 3 $\times$ R$_{200}$ from the cluster centre and the cluster redshift, as defined in equation~\ref{eqn:deltv}. The figure displays the full range of $\Delta$v within [-5000:5000] km s$^{-1}$. In the subsequent analysis, we chose to focus on the range of [-2000:2000] km s$^{-1}$ around the cluster as indicated by the dashed lines.} 
\label{fig:deltV}
\end{figure}

Figure~\ref{fig:Zhist} shows the redshift distribution of these MgII quasar absorber matches in orange while the quasar spectra with foreground clusters
are shown in blue, and the background quasars in green. We use the publicly available catalogue to estimate the velocity difference, $\Delta v$, between the absorption redshift $(z_{\rm abs})$ and the cluster redshift $(z_{\rm cluster})$. We express $\Delta v$ as follows:


\begin{equation}
    \Delta v= \frac{\Delta z \times c}{1 + z_{\rm cluster}},
    \label{eqn:deltv}
\end{equation}
where $c$ is the speed of light in km/s and $z_{\rm cluster}$ is the redshift of the clusters for the individual cluster--quasar pairs. These velocity differences can be very large, because the absorption redshifts are not necessarily close to the clusters. Figure~\ref{fig:deltV} displays the full range of  $\Delta$v within [-5000:5000] km s$^{-1}$. Due to 
the high velocity dispersion expected in galaxy clusters \citep{Girardi93}, we chose to focus on the range of [-2000:2000] km s$^{-1}$ around the clusters. Figure~\ref{fig:deltV}  indicates that a lower cut, namely [-1000:1000] km s$^{-1}$, would make little difference to the sample. In total, we found 32 absorbers associated with the clusters in our sample, resulting in an overall incidence of 32/16\,224 $\approx$  0.2\% where 16\,224 refers to the total number of quasar spectra studied (see Table \ref{tab:numberspectra}).

\section{Analysis}
\label{section:analsys}

\subsection{Normalising the background quasar spectra}
\label{subsection:normalisation}
\begin{figure*}
\centering
\includegraphics[width=0.75\textwidth]{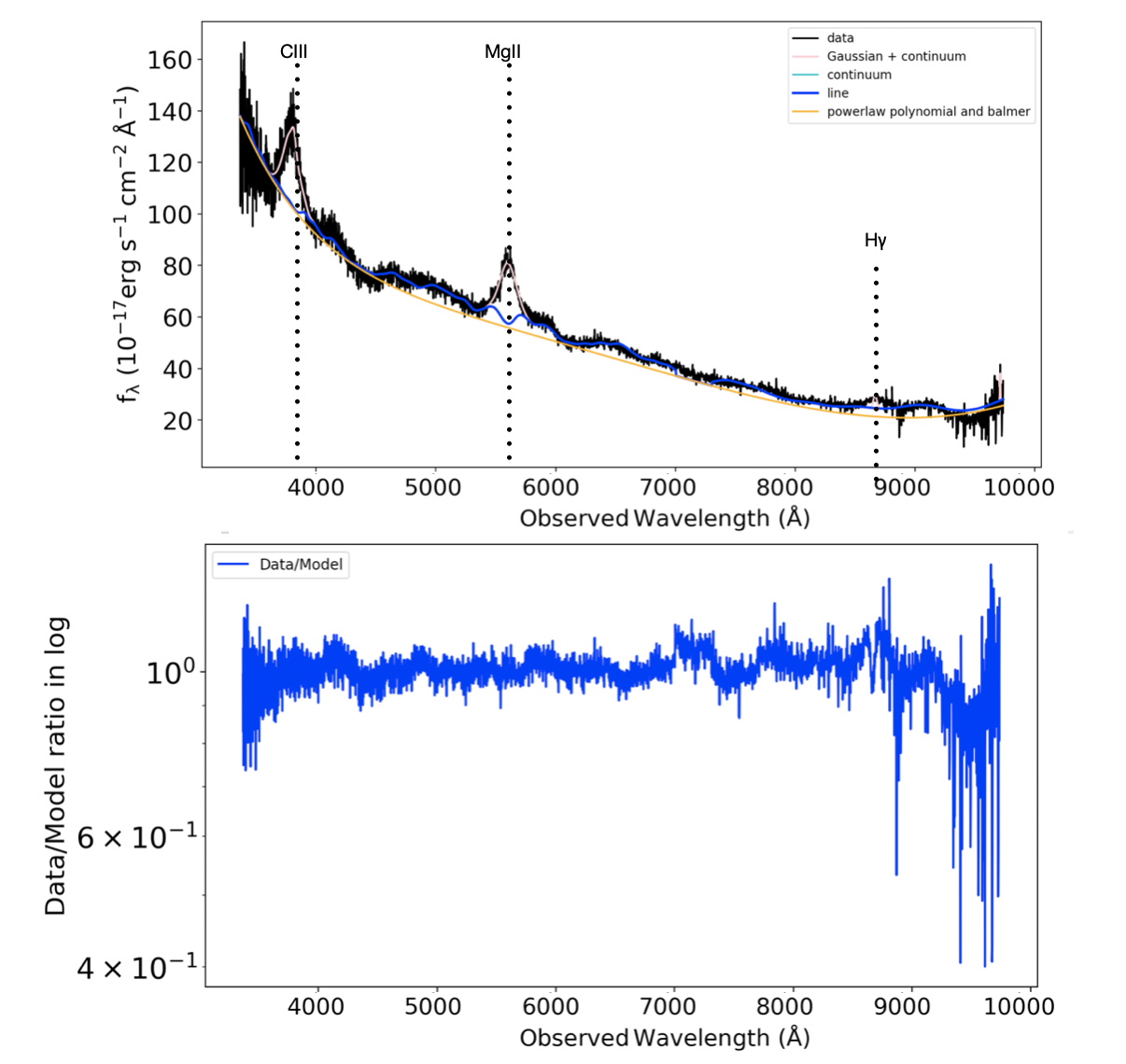}
 \caption{Example observed-frame SDSS quasar spectrum and data/continuum model ratio spectra after removing the features of quasar emission and converting to unit average flux density with the package \textsc{PyQSOfit} \citep{Guo18}. On top, we show the original SDSS spectrum, with the multi-component fit performed with \textsc{PyQSOfit} in colour. The bottom panel displays the resulting quasar spectrum normalised to an arbitrary value of 1 by dividing the observed spectrum by the continuum.}
\label{fig:norm}
\end{figure*}

In order to retrieve the quasar absorption systems, we removed the intrinsic spectral signature of the background quasars. To this end, we modelled each of the SDSS quasar spectra with the highly flexible Python QSO fitting code, \textsc{PyQSOfit} \citep{Guo18}. This algorithm fits the spectral features, broad emission lines and continuum slope of the quasars. Figure~\ref{fig:norm} displays an example spectrum showing both the quasar continuum fit and the ratio between the original data and the fitted continuum, resulting in a normalised spectrum. For some objects, the emission lines cannot be perfectly modelled and after normalisation, some residuals remain. We note in particular the noise of the SDSS spectra increases at each end of the wavelength window, resulting in noisy edges in the normalised spectra. Nonetheless, the method proved very efficient to remove the features of quasar emission.

\subsection{Stacking background quasar spectra at the position of foreground clusters}

In order to detect the weakest tracers of the cold gas, we take advantage of the large number of spectra available, and stack the background quasar spectra at the redshift of the foreground clusters. We first shift the quasar spectra to the cluster rest frame. We use the fact that all the SDSS spectra have a constant log step of 0.0001 (approximately 69 km s$^{-1}$) to avoid interpolating between pixels. We note here that for this specific cluster sample \cite{Clerc2020, Kirkpatrick21} indicate that the typical statistical uncertainty on cluster redshifts is $\Delta z/(1 + z) = 6 \times 10^{-4}$, which is about 180 km/s. This mean value does not reflect the distribution of the cluster redshift uncertainties, which may depart from a Gaussian profile. Following the method of \cite{Fresco20}, we performed a median stack in the velocity region that can be attributed to the foreground cluster within a window of [-2000:2000] km s$^{-1}$. After this process, a Gaussian smoothing of 1 pixel was applied using \textsc{gaussian smooth} from \textsc{specutils}.

In this analysis, we stacked two independent sets of quasar spectra. First, we stacked quasar spectra with known MgII absorbers that happen to be close in velocity space to a foreground cluster. We refer to this sample as the {MgII-selected sample}; this sample is used as a test for our methodology and to check whether or not the stacked spectra recover the features even with the large velocity dispersions inherent to massive galaxy clusters, which will likely induce velocity offsets at the impact parameter where the quasar's line-of-sight pierces the cold gas traced by MgII. Table~\ref{Numspec_MgII} summarises the (small) number of available spectra in the MgII-selected sample and the corresponding S/N of the resulting stacked spectrum. 

Second, we `blindly' (i.e. without prior knowledge of the presence of MgII absorbers) stack all the quasar spectra that have a foreground cluster; these correspond to the so-called {blind sample} and are described in Table~\ref{Numspec_MgII}. 

\subsection{Measuring equivalent widths and column densities}

\begin{figure}
\centering
\includegraphics[width=1\columnwidth]{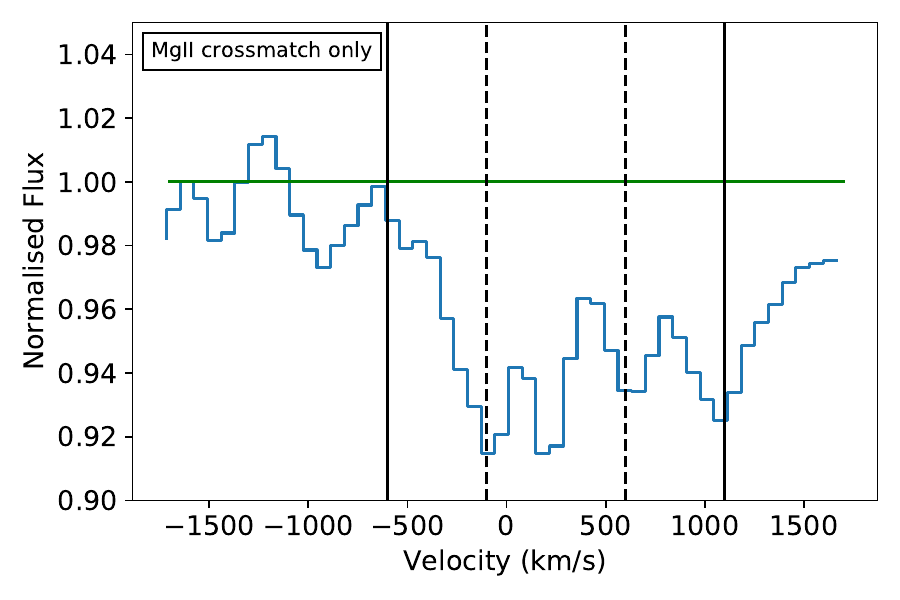}
\caption{Stack of quasar spectra with a sightline within 3 x R$_{200}$ of the centre of a cluster, and with known MgII absorbers from the MgII absorption catalogue of \cite{Anand21} satisfying $|dv|<2000$ km/s. The figures display the normalised flux in velocity space. The first dashed line at -100km/s is fixed at the location of the minimum flux within [-2000:2000]km/s of the redshift of the known MgII absorber. The second dashed line is placed at +750km/s from the first dashed line, which corresponds to the separation of the two MgII lines in the doublet. The solid vertical lines are fixed at -500km/s from 2976 and +500km/s from 2803 line. The green line shows the fitted continuum used for the EW measurement.}
\label{fig:mgsample}
\end{figure}

We first search blindly ---that is without prior knowledge of the presence of any MgII in the 
quasar spectrum--- for the minimum flux within the full velocity range [-2000:2000] km/s around the MgII 2796 \AA\ line. The wavelength of the minimum flux is set to the MgII 2796 \AA\, line. We then use the precise 750 km s$^{-1}$ velocity separation between the MgII lines of the doublet to materialise the position of MgII 2803 \AA. To quantify the amount of cold gas, we measure the EW of the MgII lines. By focusing on the MgII 2796 \AA\ line, we obtain a conservative measurement of the strength of the MgII doublet. The observed spectra can be seen in Figures ~\ref{fig:mgsample} and \ref{fig:fullstack} for the MgII-sample and blind sample, respectively. The EW$_{2796}$ is measured by integrating the area of the spectrum over the wavelength range (i.e. summing the absorbed flux values per pixel) defined by vertical lines around our first line of the doublet, which we fixed at the MgII 2796 \,\AA\, line -500km/s blueward and +375km/s redward. The continuum level is defined as the observed spectrum divided by the fitted continuum as previously described in section~\ref{subsection:normalisation}. We then measured the second line of the MgII doublet, the MgII 2803 \AA\ line, which is located at +750km/s redward. The same method is applied to the MgII 2803 \AA\ line, which is integrated from -375km/s (blueward) to +500km/s (redward) of the rest-frame wavelength. These measures result in estimates of the strength of the detected MgII absorbers. We made use of the \textsc{Linetools} software package \citep{Prochaska17}, which calculates the EW$_{2796}$ in observed wavelength space. 
We then converted the observed-frame EW$_{2796}$ to rest-frame EW, following the usual relation:

\begin{equation}
    \mathrm{EW_{rest}=\frac{EW_{obs}}{(1 + \langle z_{cluster} \rangle)}}
,\end{equation}

where the redshift $\langle\mathrm{z_{cluster}}\rangle$ is the average cluster redshift of the MgII selected sample as listed in Table~\ref{Numspec_MgII}. The estimates of the uncertainties are calculated from the mock spectra, as described in section~\ref{Subsection:Simulated spectra}.

We then computed the MgII column density according to the linear relation between the EW and the column density using the following equation from \cite{Zhu14}: 

\begin{equation}
    \mathrm{N[MgII]= 1.13 \times 10^{20} \bigg(\frac{EW_{\rm rest}} {f_{\rm osc} \lambda^{2}_{r\rm est}}}\bigg)\, \rm cm^{-2}
    \label{eqn:colden}
,\end{equation}

where $\lambda_{\rm rest}$ is the rest wavelength of each of the MgII lines of the doublet, $\mathrm{\lambda_{rest}=2796}$\,\AA\ and $\mathrm{\lambda_{rest}=2803}$\,\AA. The corresponding oscillator strengths are $\mathrm{f_{osc}=0.615}$ and $\mathrm{f_{osc}=0.306}$. For saturated absorbers with EW$_{2796}$$ > 0.15 $\,\AA, we derive a lower limit on the MgII column density (see Table~\ref{Numspec_MgII}). 

\section{Results}

\subsection{The MgII-selected sample}
\label{sec:MgII}

We used the method described in the previous section to compute the EW of the detected MgII feature in the MgII-selected stack. The corresponding measurements are listed in Table~\ref{Numspec_MgII} and the stacked spectrum is displayed in Fig~\ref{fig:mgsample}. The detection of the MgII doublet in the stack validates the methodology by showing the very strong absorption of the MgII doublet present at the redshift of the clusters. Also, despite the high-velocity dispersions within the clusters,  this detection demonstrates that the cold 10${^4}$K gas can be traced by MgII absorbers observed in the spectrum of background quasars. The detected lines are strong and show indications of saturation. 
Given the saturated EWs in the range 0.15 $<$ EW$_{2796}$ $<$ 2\AA\ \citep{Churchill00}, we calculate a lower limit for the column density using equation~\ref{eqn:colden}. The resulting EW$_{2796}$ is EW$_{2796}$=0.35$\pm0.015$ \AA\ for the MgII 2796\AA\ line, corresponding to a column density of log [N(MgII)/cm$^{-2}$]=$12.92$. The bluest line of the MgII doublet, 2796\AA, is located at -100km/s from the cluster redshift, but we note that this slight offset is eight times smaller than the typical velocity dispersion expected in galaxy clusters \citep[$\sim$800 km/s,][]{Kirkpatrick21}. This means that the expected shift from the velocity dispersion alone likely explains the small offset observed here.

To remove the effects inherent to the a priori knowledge of MgII, we also compute the {frequency-weighted} EWs and column densities as follows: $\mathrm{EW = EW_{rest} f_{c}}$, where the covering fraction, f$_c$, is defined as the number of spectra with known absorbers over the total number of spectra available for each bin of mass and angular separation. These values are listed in  Table~\ref{Numspec_MgII}.

\subsection{The blind sample}

\begin{figure}
\centering
\includegraphics[width=\columnwidth]{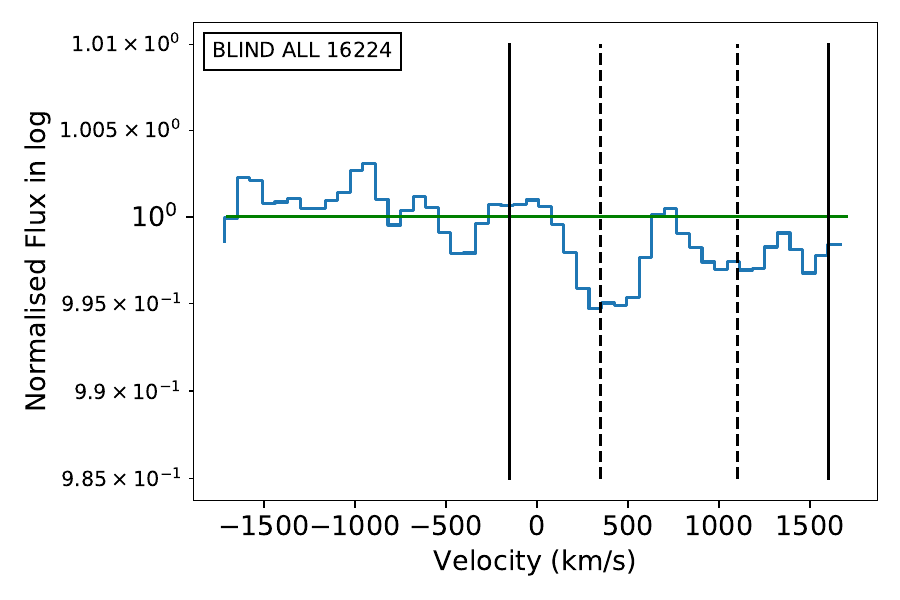}
\caption{Blind sample stack of all the background quasar spectra available at the cluster redshift, showing the normalised flux in log in velocity space. This sample comprises a total of 16\,224 spectra and leads to an S/N of the stack of 195. The detection has a rest EW of EW$_{2796}$=0.056$\pm$0.015~\AA\ (3.7$\sigma$ significance), corresponding to a column density of log [N(MgII)/cm$^{-2}]=12.12\pm0.1$. The green line shows the fitted continuum used for the
EW measurement.}
\label{fig:fullstack}
\end{figure}

 Figure~\ref{fig:fullstack} displays the stack of all the spectra, with a total of 16\,224 cluster--quasar pairs. This stack shows an absorption of the MgII doublet at the redshift of the cluster centre, with a rest EW for MgII $\lambda 2796$\AA\ of EW$_{2796}$=$0.056 \pm 0.015$\AA\ (3.7$\sigma$ significance), and a column density of log [N(MgII)/cm$^{-2}$]=$12.12\pm0.1$.  The mean mass of clusters contributing to the stacked spectra is $9.7\times 10^{14}$ M$\odot$ (M$_{500}$), and a mean angular separation of the cluster--quasar pairs is about 2 $\times$ R$_{200}$. We note that the absorption is redshifted from the cluster centre by 350 km s$^{-1}$ from the absorption line centre. This shift might be related to the presence of inflows and/or outflows, although it is challenging to determine this from absorption line studies alone because the orientation of the clusters and sightlines might complicate the interpretation. The uncertainty in the measurements was calculated by randomising the redshift of the clusters and repeating the stacking 500 times, as described in section~\ref{Subsection:Simulated spectra}.
 
  The detection is broad, ranging over 1\,000 km/s in width for each line. This is likely the result of the offset in velocities in the different clusters along the background sightlines. In addition, we computed the stack of the blind sample after removing the known MgII absorbers. We find that the absorption features remain largely unchanged, which further supports the idea that the detected absorption feature is not solely the result of a few known strong absorption systems. 

\begin{table*}
\setlength\tabcolsep{2pt}
\footnotesize
\caption{Properties of the two stacks performed. }
\begin{threeparttable}
\begin{adjustbox}{width=2\columnwidth,center} 
\begin{tabular}{ccccccccccc}
\hline \hline
 & \# Quasar & Mean & S/N & EW MgII & EW MgII  & MgII & Col Den 2796 &  Col Den 2803 & Frequency\\[0.1cm]
    Sample &  Spectra & MgII & MgII & 2796 & 2803 &  Doublet  &  log N(MgII) & log N(MgII) & Weighted   2796 \\[0.1cm]
 & & Redshift & sample  & \AA &  \AA &Ratio & $\mathrm{cm^{-2}}$ &  $\mathrm{cm^{-2}}$ & log N(MgII) $\mathrm{cm^{-2}}$\\[0.1cm]
 \hline  MgII selected & 32/16224=0.2\% & 0.51 & 73 & $0.35\pm0.015$ & 0.19$\pm0.015$ & 1.8 & $\geq$12.92 & 12.95$\pm0.1$ & $\geq10.22$ \\[0.11cm]
  \hline
    Blind & 16224 =100\% & 0.45 & 195 & $0.056\pm0.015 $ & $ 0.024\pm0.015$  & 2.3 & $12.12\pm0.1$ & $12.05\pm0.1$ & $12.12\pm0.1$\\[0.11cm]
  \hline \hline
\label{Numspec_MgII}
\end{tabular}
\end{adjustbox}
\footnotesize{Table \ref{Numspec_MgII} lists the number of background quasar spectra stacked. The column of the number of quasar spectra corresponds to the fraction of quasars over the entire number of spectra with its representative percentage. The column density for the MgII-selected sample is frequency-weighted, considering the number of spectra used. Possible saturation (at EW$_{2796}$ $>$ 0.15 \AA) leads to a lower limit on the column density of the MgII-selected sample. MgII absorption is tentatively detected in the blind stack. This tentative detection provides a measure of the column density, which relates to the amount of cold gas in galaxy clusters.}
\end{threeparttable}
\end{table*}

\subsection{Mock Sloan spectra with simulated MgII absorbers}
\label{Subsection:Simulated spectra}

In order to quantify the systematic uncertainty of the method and derive an error on the EW measurements, we use 22\,000 S/N=200 mock Sloan stacks representative of the S/N achieved with our observed stack. The synthetic spectra are convolved with the line spread function (LSF) of the  the SDSS instrument to achieve the same spectral resolution as the observations. Similarly, noise properties typical of Sloan spectra are included. We used Gaussian noise with a spread similar to Sloan's typical error, including increased noise towards each end of the spectrum. We additionally inserted an MgII absorber doublet from the TNG50 cosmological magnetohydrodynamical simulations. To compute MgII, we take the total magnesium mass per cell as tracked during the simulation, and use \textsc{CLOUDY} \citep{cloudy} to calculate the ionisation state assuming both collisional and photo-ionisation following the modelling approach of \cite{Nelson20}. We note that the \textsc{CLOUDY} modelling indicates that at densities of $<$ 10$^{-2.5}$ cm$^{-3}$, the photo-ionisation dominates, but overall both photo-ionisation and collisional ionisation processes are expected to play a role at densities typical of the ICM \citep{Peterson01}. We then ray-trace through the simulated gas distribution to create synthetic absorption spectra akin to those in real observations \citep[][Nelson et al. in prep.]
{Szakacs23}. This is similar in its rationale to several other techniques for creating absorption spectra from hydrodynamical simulations, such as \textsc{specwizard} \citep{Theuns98, Shaye03}, \textsc{Trident} \citep{Hummels17}, and \textsc{pygad} \citep{pygad21}. The absorbers are inserted at a given wavelength position but with different EWs, ranging from 0.05 to 5 \AA. We used these mock spectra to run the same search and measurement algorithm as that used for analysing the observations. Our results show that in 22\,000 mock MgII stacks, only 3\% of the detected absorption as found by the minimum-flux pixel approach is not associated with a simulated input MgII line. This means that the code finds the minimum flux outside of the [-2000:2000]km/s  area where the MgII doublet should be located at a given redshift.

We then measured the EW$_{2796}$ of the simulated spectra with the method used to analyse the observations. 
We find that the EW measurements from our analysis are consistent with the input EW from the simulations. The analysis also indicates that the mean difference between the measured values and the input EW is 0.013 \AA\, for EWs typical of the one we measure in the blind stack. This is similar to our estimated error in the blind stack of 0.015\AA.

\begin{figure}
\centering
\includegraphics[width=1\columnwidth]{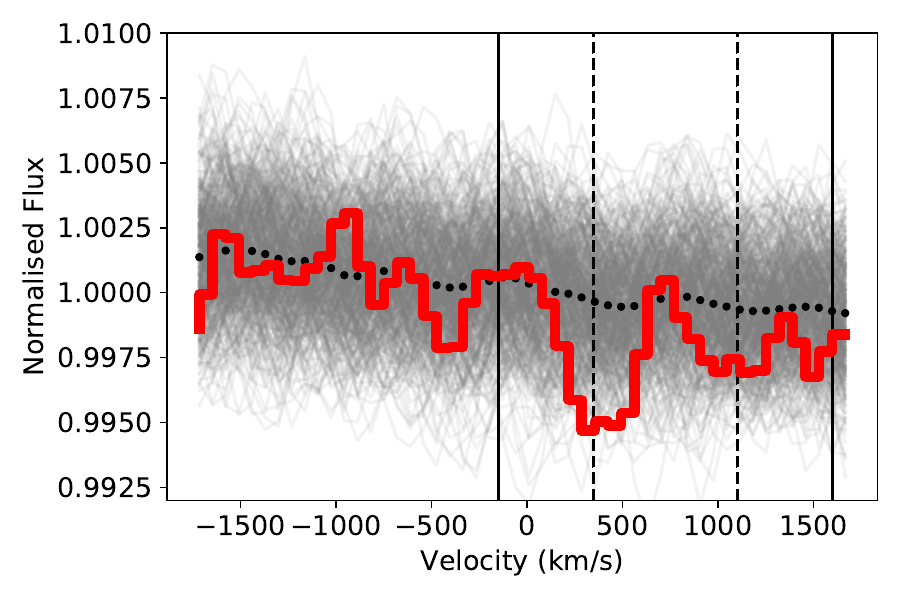}
\caption{Plot of 500 random stacks from the bootstrapping in grey. The stacks are from the same sample of 16\,224 cluster--quasar pairs stacked at randomised redshifts from the full redshift distribution. The overlayed black dotted line represents the median flux per velocity bin of the 500 random stacks. On top, we over-plot in red the original blind stack from Figure~\ref{fig:fullstack} to illustrate the difference. }
\label{fig:bootstrapping}
\end{figure}
  
\begin{figure}
\centering
\includegraphics[width=1\columnwidth]{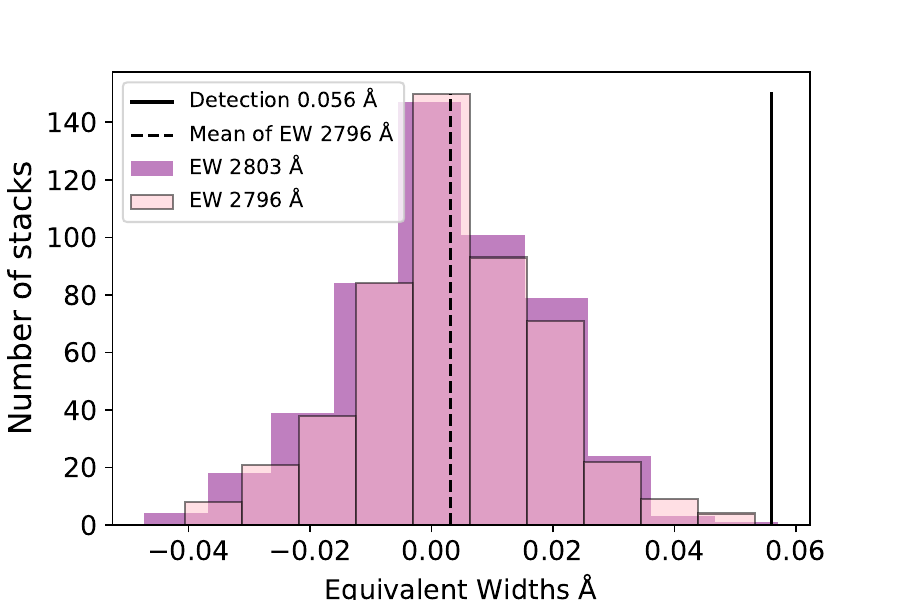}
\caption{Histogram of the measurements of the EWs of the 500 random stacks from bootstrapping for the MgII 2793 and 2803 \AA\ lines stacked at random redshifts from the cluster redshift distribution. We mark with a vertical solid line the position of our measurement of EW$_{2796}$ corresponding to 0.056$\pm0.015$ \AA, and with a vertical dashed line the mean of EW$_{2796}$ of 0.003 \AA. We measure a standard deviation from the distribution of EW$_{2796}$ of 0.015\AA, which, considering our tentative detection, gives us a significance of 3.7$\sigma$. }
\label{fig:ewhist}
\end{figure}

\subsection{Uncertainty assessment from bootstrapping}
\label{Subsection:Simulated spectra}

In addition to using the mock spectra with simulated absorbers to quantify the possible systematic errors of our method, we also performed a non-parametric bootstrapping experiment by stacking the full sample of cluster--quasar pairs ---at cluster redshifts that have been randomised--- 500 times. We then repeated the  measurement process mentioned in section~\ref{section:analsys}, where we first find the minimum in the flux, and then measure the EW of the possible absorption feature. We do a binning of the 500 stacks for each flux point, and calculate the median of each bin, obtaining one final median array shown as the black dotted line in Figure~\ref{fig:bootstrapping}, where the grey lines in the background represent each random stack, and the overlayed black dotted line represents the medium flux per velocity bin of the 500 random stacks. We further overplot in red the blind stack from Figure~\ref{fig:fullstack}. The results show that the stacks at random redshifts do not display absorption features at the level of the blind stack shown in red. This demonstrates that the detection is significant. Indeed, the measurement of the EW of the blind stack is higher than the distribution of the measured EW of the absorption features of the random stacks as seen in Figure~\ref{fig:ewhist}. As expected, the distribution peaks around zero, because we do not expect any absorption feature at the random redshift we are considering. 
 
  Assuming a normal distribution, we get a standard deviation from the EW distribution of 0.015\AA. Taking this as our measurement uncertainty gives us a 3.7$\sigma$ significance for the MgII 2796\AA\ absorption feature.


\section{Discussion}

\subsection{Evidence of cold gas in X-ray-selected clusters}

The detection of strong MgII absorption in the MgII-selected sample validates our approach of stacking many medium-S/N background quasar spectra to increase the sensitivity to cold gas in clusters. This detection further demonstrates that, despite the high velocity dispersions within the clusters, the 10${^4}$K gas can be traced by the MgII absorbers observed in the spectrum of background quasars. Our findings based on the blind stack of the full sample also indicate the presence of some cold 10$^4$K gas traced by MgII in the intracluster environment. We note that the CGM has a rather loose definition, and so whether these MgII absorbers are associated with the circumgalactic gas of individual galaxy members or with the intracluster gas is somewhat subjective. More important in this work is the total amount of 10$^4$ K cold gas that is being detected in the dense cluster regions. These observational results are in line with expectation from various numerical \citep{Sharma12, Mccourt12} and hydrodynamical simulations \citep{Pillepich18, Nelson18, Tremmel19, Butsky19}, which show that the interactions of supermassive black holes and large-scale atmospheres of massive clusters cause the CGM to cool down rapidly and homogeneously \citep{Donahue22}. The low entropy in groups of galaxies and galaxy clusters explains how the cold gas filaments can condense inside a hot halo, where local thermal instability makes it possible for cool gas blobs to survive the hot medium \citep{Sharma12, Mccourt12}. 

To put our results into perspective, we plot in Figure~\ref{fig:other_works} the EW (left y-axis) and column density (right y-axis) of MgII-absorbing gas as a function of the foreground cluster mass (right panel) and the projected distance expressed in kiloparsecs (kpc) (left panel) in comparison with similar studies, which we discuss further below.


\subsection{Cold gas in galaxy clusters}

 \cite{Lopez} were the first to look for cold gas in foreground galaxy clusters using background quasar spectra. They based their study on a sample of 442 cluster--quasar pairs. To this end, the authors made use of the third data release of the SDSS with high-redshift cluster or group candidates from the
Red-Sequence Cluster Survey \citep{Gladders05}. \cite{Lopez} found that there is proportionally less cold gas in more massive clusters than in low-mass systems when using models of galaxy counts. This refers to the relation between overdensities of MgII absorbers in clusters with much denser galaxy environments. When considering the stellar baryon fraction, studies still disagree on the exact slope of the stellar and total baryon fraction as a function of cluster halo mass \citep{Gonzaleznov13}.  Nevertheless, the \cite{Lopez} results are in agreement with our findings: the most massive structures of their sample (M $> 10^{14}$ M$_{\odot}$) have an overdensity of absorbers with EW$_{2796}$$>$ 1\AA\ that is  twice higher than for moderate mass clusters (M $\sim 2\times10^{13}$ M$_{\odot}$). The effect is even more pronounced at smaller distances (d$<$ 1Mpc) from the cluster centre. At the same time, \cite{Lopez} point out that the more massive clusters contain five times more galaxies than the less massive ones. While searching for this cold gas through MgII absorption lines, these latter authors find that a subsample of their massive clusters yield a stronger and more significant signal. Recently, \cite{Lee} performed a  cluster--quasar cross-correlation with SDSS DR14 quasars and redMaPPer clusters, with a total of 82\,000 cluster--quasar pairs. Although there was no stacking of quasar spectra in this latter work, the authors report that the MgII absorber detection rate per quasar is 2.70$\pm$0.66 times higher inside the clusters than outside them. This shows that Mg II absorbers are abundant in clusters compared to the field.\\

\begin{figure*}
\centering
\includegraphics[width=\columnwidth]{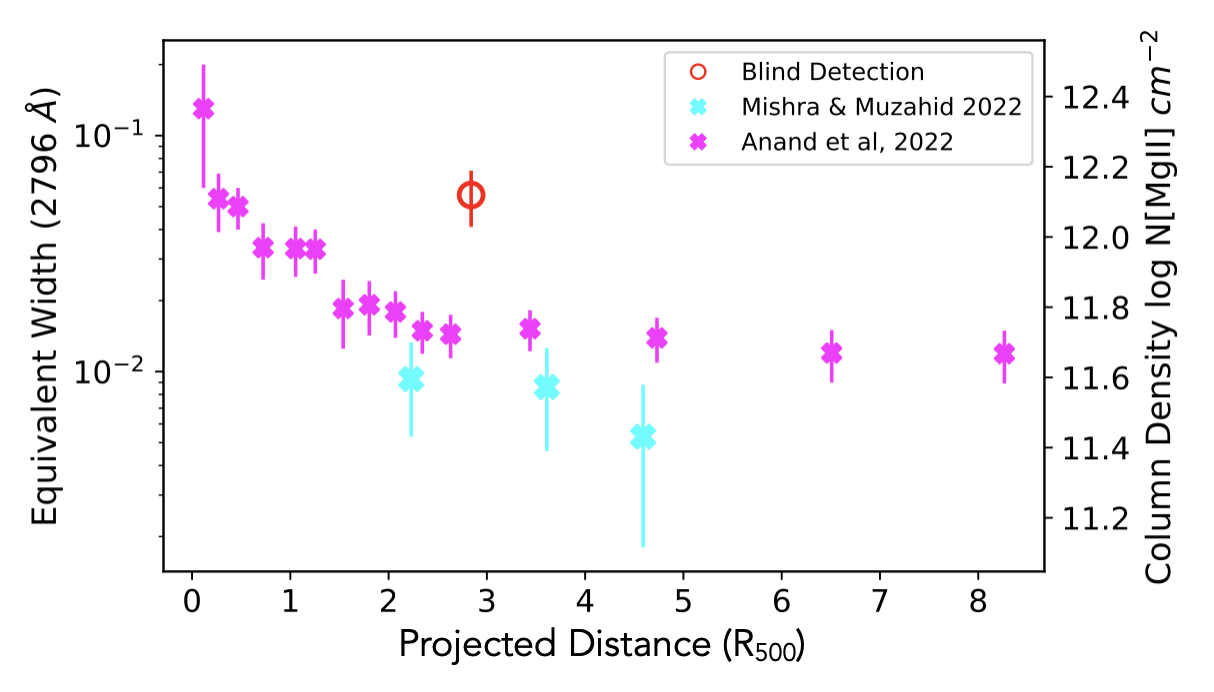}
\includegraphics[width=1\columnwidth]{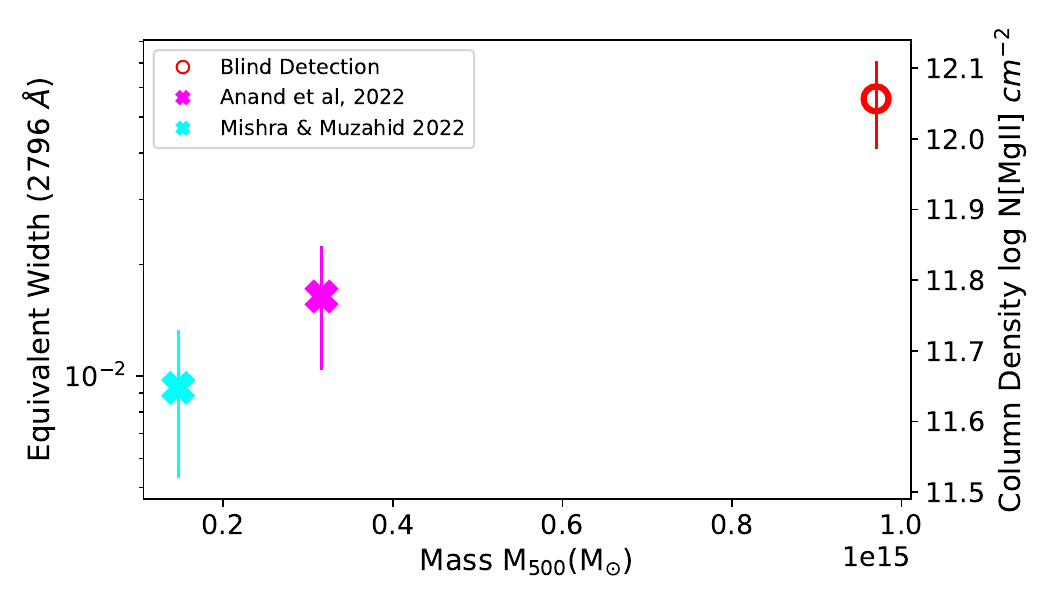}
\caption{Equivalent width and column density from the blind sample compared with results from the literature \protect\citep{Anand21, Mishra}. The left panel displays the detection in the blind sample as a function of projected distance (in units of R$_{500}$), with the detection in our blind stack depicted as a red circle. In the right panel, we plot the mean of the two points of \protect\cite{Anand21} and the point of \protect\cite{Mishra} closest to our results in projected distance space (see left panel). This second figure highlights the difference in mass range between the samples. }
\label{fig:other_works}
\end{figure*}

\begin{figure}
\centering
\includegraphics[width=\columnwidth]{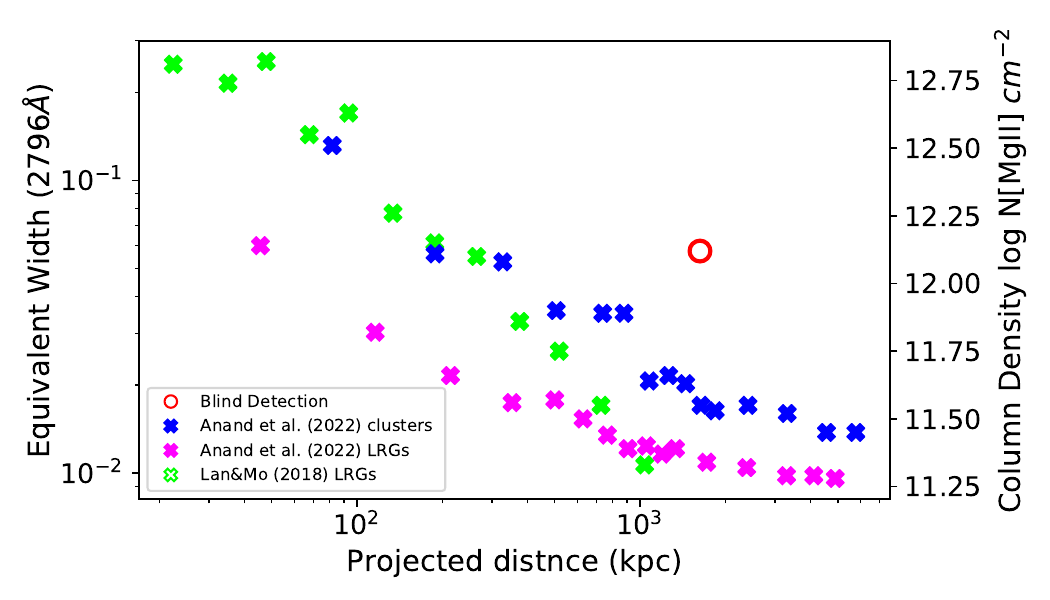}
\caption{Comparison between our results and measurements of column density and EW of MgII absorption in LRGs from \protect\cite{Lan&Mo18}, LRGs and cluster samples from \protect\cite{Anand22}, and our tentative detection from the blind sample, as a function of projected distance.}
\label{fig:LRG}
\end{figure}

Using a photometric sample of galaxy clusters
identified in the Legacy Imaging Survey of Dark Energy Survey Instrument (DESI) by \citet{Zou21}, a follow-up study led by \cite{Anand22} cross-correlates the MgII absorption catalogue \citep{Anand21} with cluster and luminous red
galaxies (from SDSS BOSS survey) with halo masses in the range between $10^{13.8}$ and $10^{14.8}$ M$_\odot$. 
Furthermore, \cite{Mishra} cross-match the SDSS cluster catalogue of \cite{Wen12} and the SDSS DR16 quasar catalogue of \cite{SDSSLyke}. The median impact parameter of the clusters from the quasar sightlines is 2.4 Mpc (median 3.6 R$_{500}$).  \cite{Mishra} were the first authors to stack the background quasars to increase the sensitivity of the experiment; in their work, they measured the total EW and assumed that the observed MgII line falls on the linear part of the curve of growth, meaning that the EW$_{2796}$ is two-thirds of the total EW.\\

Figure~\ref{fig:other_works} displays our results together with the EWs from the above surveys. The left panel of Figure~\ref{fig:other_works} displays the EW from the blind sample in our work compared to the values reported by \cite{Anand22} and \cite{Mishra} at different values of projected distance. The right panel of Figure~\ref{fig:other_works} shows the EWs as a function of cluster mass. Clearly, the works of \cite{Anand22} and \cite{Mishra} probe a lower mass range than our study. We compare the median of the two closest numbers in projected distance from \cite{Anand22}, and the closest point from \cite{Mishra} with our tentative detection (red circle). Both \cite{Mishra} and \cite{Anand22} also report a clear trend of decreasing EW$_{2796}$ with increased projected distance. Despite the differences between the approaches, the reported EW is in line with results from \cite{Anand22}. The cluster masses in our study range from M$_{500}=1 \times 10^{14}$ to $8.4 \times 10^{15}$, with an EW$_{2796}$ of 0.056$\pm0.015$ \AA\ of MgII 2796 line.  Using three times the number of quasar cluster pairs, the rest EW$_{2796}$ from \cite{Mishra} remain one order of magnitude smaller than our measured rest EWs from both our MgII-selected sample and our blind sample, as shown in the lower panel of Figure~\ref{fig:other_works}.\\

Our higher EW measurements and limits could be associated with the higher mass in our sample, as shown in Figure~\ref{fig:other_works}. Indeed, \citet{Sharma12, Mccourt12, Donahue22} propose that although cooling is negligible near the virial radius, it becomes more important at higher densities and smaller radii, which is in line with these observations. Additionally, complementary observational studies have analysed the incidence of MgII absorbers in the context of isolated and group environments. Specifically, MgII metal line emission studies from galaxies at z$\sim$1 find extended MgII emission in a large blind galaxy survey in the Muse Analysis of Gas around Galaxies or MAGG \citep{Dutta20}. Their results show that the MgII is associated with multiple galaxies, and their measurements of the MgII emission flux in these groups of galaxies are on average five times stronger than in isolated galaxies. Their findings therefore favour the scenario where the hydrodynamic interactions among group members are the primary reason for the increased strength of MgII lines. Similarly, observational studies of the cold gas in the CGM at higher redshifts (z$\sim$2) by \citet{Nielsen20} reveal a contrast, in that these galaxies are actively forming at cosmic noon, where the CGM is less massive than at lower redshifts. The measurements of MgII EW are found to be larger (EW$_{2796}$$\geq$ 0.5\AA) in the lower-redshift (z$\sim$1) studies (i.e. at higher mass) from the MusE GAs FLOw and Wind survey or MEGAFLOW \citep{Schroetter16, Schroetter19} than at z$\sim$2. \\
\\


Figure~\ref{fig:LRG} contrasts our results for EW$_{2796}$ with observations of lower-mass structures, including photometric redshift clusters from the legacy imaging survey of the
Dark Energy Survey \citep{Anand22}, luminous red galaxies (LRGs) from SDSS DR16 \citep{Anand22}, and SDSS DR14 \citep{Lan&Mo18} as a function of projected distance. The LRGs with typical masses of $\sim 10^{11.2}$ M$_\odot$ from \cite{Anand22} show a smaller average MgII EW than the clusters. Although these values seem to converge at large impact parameters, they report this difference as being due to the larger halo masses and denser environments in the cluster sample. On the other hand, the LRGs from \cite{Lan&Mo18} show larger EW$_{2796}$ at lower impact parameter, and a steeper decline at higher projected distance compared to clusters from the legacy imaging survey of the Dark Energy
Survey and the LRGs from \cite{Anand22}. 
 We may again attribute the difference in EW$_{2796}$ we report in this study to the mass differences, because our sample of clusters has a higher mean mass than the samples used in these previous works (see lower panel of Figure \ref{fig:other_works}). \\

While there still exists disagreement in the trend of total baryon fraction as a function of cluster mass \citep{Gonzaleznov13, Lagan13}, the difference in mass between all the previously mentioned studies and our work, in connection with the higher measurements from EW from our higher-cluster-mass sample, reveals a trend: increased baryon fraction at higher cluster masses. These differences in gas measurements for halos of different mass can also be seen when comparing galaxies versus galaxy clusters as shown in Figure~\ref{fig:LRG}. Similarly, \cite{Anand22} perform a direct comparison of LRGs with legacy imaging
survey clusters, finding consistently higher MgII EWs in the cluster sample than in the LRGs, where the clusters are between two and three times more massive than LRGs within R$_{500}$.



\subsection{Comparison with simulations}

\cite{Butsky19} used the RomulusC simulations \citep{Tremmel19} to probe the nature of the multi-phase cool--warm (10$^4$ $<$ T $<$ 10$^6$K) gas in and around a galaxy cluster of  mass 10$^{14}$ M$_{\odot}$. Their study makes predictions for the covering fractions of key absorption-line tracers, both in the ICM and CGM of cluster galaxies using synthetic spectra. The authors find there is a significant quantity of multi-phase gas in the cool (10$^{4-5}$K) gas at all clustocentric radii. The results from \cite{Butsky19} indicate that the column density of all ions declines from the cluster centre out to 1 Mpc, but remains relatively flat towards the edge of the halo. In a more recent set of CGM simulations around massive galaxies and groups of galaxies dubbed Romulus, \citet{Saeedzadeh23} expand their previous simulations, showing that the presence of cold gas in the CGM can be described as filaments of inflowing cooling gas as well as gaseous tails from possible satellites. Furthermore, condensation patches can also originate from density perturbations cooling rapidly, which provides further theoretical support to the observations of cold gas in clusters. \\

\begin{figure}
\centering
\includegraphics[width=\columnwidth]{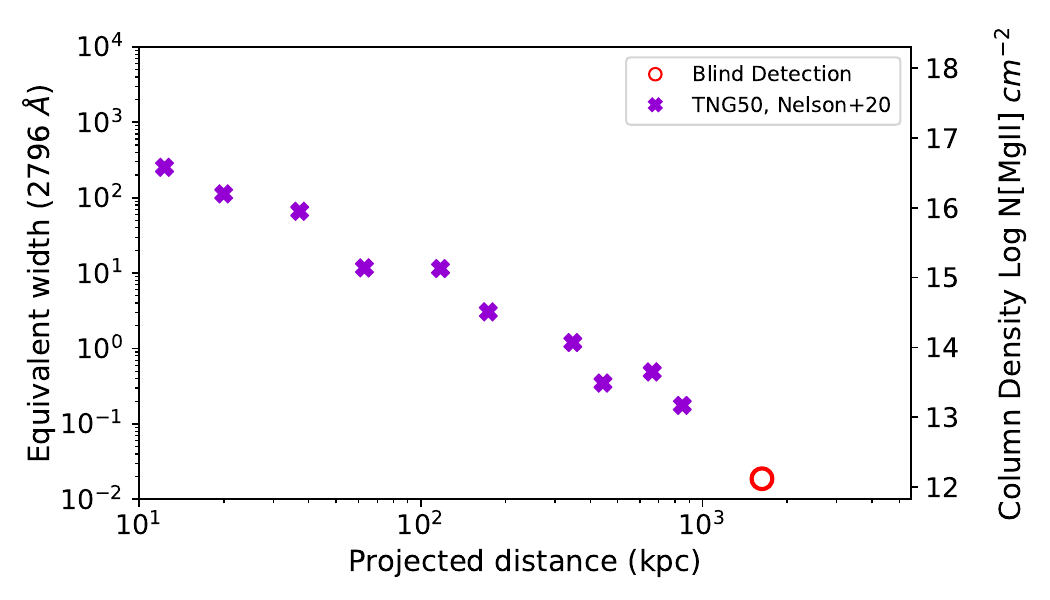}
\caption{Comparison between our blind stack with hydrodynamical cosmological simulations from TNG50 \citep{Nelson20}. The shape and colour of the symbols are as in previous figures. The purple symbols display predictions of the MgII column density in TNG50 as a function of projected distance in kpc. We stress that there are large mass differences between the models and the data: the TNG50 highest mass bin ($\sim 10^{13.5}$ M$_{\odot}$) displayed in the figure is smaller than our smallest mass cluster ($3.28 \times 10^{14}$ M$_{\odot}$). Considering this mass difference, we note that there is  likely an excess of MgII-absorbing cold gas in the TNG50 simulations.}
\label{fig:TNG}
\end{figure}

In Figure~\ref{fig:TNG}, we compare our results with the TNG50 simulation \citep{Pillepich19, Nelson19} from the TNG suite \citep{Marinacci18, Springel18, Naiman18, Nelson18, Pillepich18}. The aim of TNG is to study the physical processes that drive galaxy formation and to investigate how galaxies evolve within large-scale structures. TNG50 is a gravomagnetohydrodynamics (MHD) cosmological simulation including a comprehensive model for galaxy formation physics \citep{Weinberger18, Pillepich18} at the highest resolution. TNG50 includes 2$\times$2160$^3$ resolution elements in a $\sim$\,50 Mpc (comoving) box. \cite{Nelson20} make predictions of the physical properties of the cold gas traced by MgII in the CGM of galaxies and groups at z = 0.5 based on these TNG-50 simulations. By identifying discrete structures of cool, MgII-rich gas, the authors found that cold gas in these massive halos is made up of thousands to tens of thousands of small (kpc), discrete clouds. The results also indicate a tendency whereby the most massive halos have the highest covering fraction, while the highest column densities are located near the halo centre, where the gas densities are greater. When specifically addressing the physics of small scales in the CGM, simulations are facing the complication of resolving these lower-density structures. This convergence issue arises from limitations in computational resources that limit the maximum resolution, and the priorities on the coding side that are usually set on denser structures. To counter this issue, \citet{Freeke19} used standard mass refinement and additional uniform spatial refinement to instead increase the resolution in the CGM of a Milky-Way-mass galaxy. Their findings demonstrate a drastic change in the radial profile of neutral hydrogen column density compared with previous simulations. The authors found both the HI covering fraction and  column density to increase. Recently, major enhanced CGM resolution cosmological simulations with the zoom-in approach using the TNG galaxy formation model have become available \citep[GIBLE,][]{Ramesh23}. The results indicate that by improving the mass resolution, the cold gas in CGM regions is better resolved. The increase in resolution in the work of \cite{Ramesh23} leads to a larger number of small clouds. While it is worth mentioning that the galactic scales simulated in these works are not directly comparable to our study, the convergence issue is relevant.

Figure~\ref{fig:TNG} shows our results in terms of the column density of MgII as a function of projected distance from the cluster. The conversion of angular size from R$_{500}$ in degrees to linear size in kpc at a given redshift was performed using cosmological parameters from \cite{Planck18}. We note that the TNG50 simulations probe significantly smaller impact parameters than our observations. Interestingly, the EW in TNG50 changes by more than three orders of magnitude from small (10 kpc) to large (1 Mpc) scales. However, it should be noted that \cite{Nelson20} provide predictions in terms of column density derived from integrating the gas density of a halo along the wavelength axis to calculate its surface density. The conversion to EWs is done in the present study using an empirically based conversion described in equation~\ref{eqn:colden}, which might be less reliable at these extreme column-density values. Equally, we stress that there are large mass differences between the models and the data: the TNG50 highest mass bin ($\sim 10^{13.5}$ M$_{\odot}$) displayed in the figure is smaller than our smallest mass cluster ($3.28 \times 10^{14}$ M$_{\odot}$). Considering this mass difference, and the observed increase in the cold column density as a function of cluster mass, we note here a possible disagreement and a likely excess of MgII-absorbing cold gas in the TNG50 simulations compared to observations. To address these findings, a more appropriate comparison would make use of the new TNG-Cluster simulations introduced in \citet{Nelson23}. Indeed, these simulations significantly increase the statistical sampling of the most massive and rarest objects in the Universe, specifically galaxy clusters evolved in cosmic time to z=0 masses of log(M$_{200c}$ / M$_{\odot}$) $>$ 14.3 - 15.4. These objects would be most relevant to the observational results presented here. Nevertheless, the analysis of the MgII gas properties in these newly released simulations is beyond the scope of the present paper. \\

\subsection{Looking forward}

The analysis presented here stems from the ambitious endeavours of the SDSS. This series of multi-object surveys, based on extremely large surveys of thousands of quasar absorbers, brought quasar studies to a new era \citep[e.g.][]{noterdaeme12, bird17, Parks18}. Such surveys advanced the field significantly because they produced homogeneous data products for well over one million low-resolution quasar spectra. In the near future, dedicated spectroscopic surveys on 4m class telescopes will provide a wealth of new low- and medium-resolution quasar spectra in extremely large numbers, notably the DESI experiment \citep{DESI16}, the WEAVE-QSO survey \citep{Weave22}, and a surveys with the 4MOST experiment \citep[including][]{Merloni19, Peroux23}. In particular, DESI is projected to obtain 3 million quasar spectra and 800\,000 MgII absorbers. In collaboration with new generations of X-ray missions, including the full eROSITA survey \citep{Merloni12}, X-ray imaging and Spectroscopy mission (XRISM; \citealp{XRISM22}), and ESA/Athena \citep{Nandra13}, the study presented here can be expanded greatly by not only increasing the number of background quasar spectra, but also increasing the spectral resolution of these data. We note that the latter is key as the EW limit of detectable absorbers scales linearly with spectral resolution, and a 1 dex increase in EW provides ten times more absorbers.

\section{Conclusions}

In this work, we used background quasar spectra from the SDSS/SPIDERS survey to explore the cold-gas content of foreground massive X-ray-selected galaxy clusters. To this end, we stacked approximately $ 16\,000$ quasar spectra in the quasar sightline. Our main results are can be summarised as follows:
\begin{itemize}

    \item From the sample of known MgII absorbers within [-2000:2000] km s$^{-1}$ of the clusters, we detect strong absorption rest EW for MgII $\lambda 2796$ of EW$_{2796}$=0.35$\pm0.015$ \AA. The uncertainty is calculated through bootstrapping. As the absorption lines are saturated (at EW$_{2796}$$>$0.15 \AA), we derive a lower limit for the column density of log [N(MgII)/cm$^{-2}$]$\geq$12.92. These results validate the technique and demonstrate that despite the high velocity dispersions within the clusters, the cold 10$^4$K gas can be traced by MgII absorbers observed in the spectra of background quasars. 
\\
    \item We then stacked quasar spectra without prior knowledge of the presence of absorption systems (so-called blind sample). We tentatively ($\approx3.7 \sigma$ significance) detect MgII absorption signal by stacking a total of 16\,224 quasar absorber--cluster pairs within [-2000:2000] km s$^{-1}$ from the clusters, with EW$_{2796}$$=0.056\pm0.015$ \AA, corresponding to a column density of log [N(MgII)/cm$^{-2}$]$=$12.12$\pm0.1$ $\mathrm{cm^{-2}}$.
\\  
    \item While we are probing a different cluster mass range, our results are in line with observational findings previously published in the literature \citep{Lee, Anand22, Mishra}. We also report an excess of MgII gas in the predictions from TNG50 hydrodynamical cosmological simulations \citep{Nelson20} compared to our work and others with lower-mass objects in their sample \citep{Anand22,  Lan&Mo18}.

\end{itemize}

Using a similar approach of stacking in upcoming surveys with increased spectral resolution and number of background spectra  \citep[DESI,][]{DESI16}, WEAVE \citep{Weave22} and 4MOST \citep{Merloni19}, combined with a new generation of X-ray facilities (most notably eROSITA \citep{Merloni12} on board SRG), will likely provide additional insight into the cold-gas content of galaxy clusters. In terms of simulations, TNG50 profiles could be extended out to an angular separation of 4000 kpc to
overlap with the observations. Further, using the new TNG-Cluster simulations \citep{Nelson23} would provide a sample of halo masses that overlap
with our observations, which would allow us to assess the importance of the reported halo mass mismatch.

\section*{Acknowledgements}
This research made use of Astropy,\footnote{http://www.astropy.org} a community-developed core Python package for Astronomy \citep{astropy18}. We thank Dylan Nelson for generating the TNG50 MgII spectra. A. Fresco gratefully acknowledges the support provided by the fellowship from Becas Carlos Antonio Lopez (BECAL), Paraguay.

\section*{Data availability}
All the data used in this work are publicly available. 

\bibliographystyle{aa}
\bibliography{biblio}

\begin{thebibliography}{119}
\expandafter\ifx\csname natexlab\endcsname\relax\def\natexlab#1{#1}\fi

\bibitem[{{Ahumada} {et~al.}(2020){Ahumada}, {Prieto}, {Almeida}, {Anders}, {Anderson}, {Andrews}, {Anguiano}, {Arcodia}, {Armengaud}, {Aubert}, {Avila}, {Avila-Reese}, {Badenes}, {Balland}, {Barger}, {Barrera-Ballesteros}, {Basu}, {Bautista}, {Beaton}, {Beers}, {Benavides}, {Bender}, {Bernardi}, {Bershady}, {Beutler}, {Bidin}, {Bird}, {Bizyaev}, {Blanc}, {Blanton}, {Boquien}, {Borissova}, {Bovy}, {Brandt}, {Brinkmann}, {Brownstein}, {Bundy}, {Bureau}, {Burgasser}, {Burtin}, {Cano-D{\'\i}az}, {Capasso}, {Cappellari}, {Carrera}, {Chabanier}, {Chaplin}, {Chapman}, {Cherinka}, {Chiappini}, {Doohyun Choi}, {Chojnowski}, {Chung}, {Clerc}, {Coffey}, {Comerford}, {Comparat}, {da Costa}, {Cousinou}, {Covey}, {Crane}, {Cunha}, {Ilha}, {Dai}, {Damsted}, {Darling}, {Davidson}, {Davies}, {Dawson}, {De}, {de la Macorra}, {De Lee}, {Queiroz}, {Deconto Machado}, {de la Torre}, {Dell'Agli}, {du Mas des Bourboux}, {Diamond-Stanic}, {Dillon}, {Donor}, {Drory}, {Duckworth}, {Dwelly}, {Ebelke}, {Eftekharzadeh}, {Davis
  Eigenbrot}, {Elsworth}, {Eracleous}, {Erfanianfar}, {Escoffier}, {Fan}, {Farr}, {Fern{\'a}ndez-Trincado}, {Feuillet}, {Finoguenov}, {Fofie}, {Fraser-McKelvie}, {Frinchaboy}, {Fromenteau}, {Fu}, {Galbany}, {Garcia}, {Garc{\'\i}a-Hern{\'a}ndez}, {Oehmichen}, {Ge}, {Maia}, {Geisler}, {Gelfand}, {Goddy}, {Gonzalez-Perez}, {Grabowski}, {Green}, {Grier}, {Guo}, {Guy}, {Harding}, {Hasselquist}, {Hawken}, {Hayes}, {Hearty}, {Hekker}, {Hogg}, {Holtzman}, {Horta}, {Hou}, {Hsieh}, {Huber}, {Hunt}, {Chitham}, {Imig}, {Jaber}, {Angel}, {Johnson}, {Jones}, {J{\"o}nsson}, {Jullo}, {Kim}, {Kinemuchi}, {Kirkpatrick}, {Kite}, {Klaene}, {Kneib}, {Kollmeier}, {Kong}, {Kounkel}, {Krishnarao}, {Lacerna}, {Lan}, {Lane}, {Law}, {Le Goff}, {Leung}, {Lewis}, {Li}, {Lian}, {Lin}, {Long}, {Longa-Pe{\~n}a}, {Lundgren}, {Lyke}, {Ted Mackereth}, {MacLeod}, {Majewski}, {Manchado}, {Maraston}, {Martini}, {Masseron}, {Masters}, {Mathur}, {McDermid}, {Merloni}, {Merrifield}, {M{\'e}sz{\'a}ros}, {Miglio}, {Minniti}, {Minsley}, {Miyaji},
  {Mohammad}, {Mosser}, {Mueller}, {Muna}, {Mu{\~n}oz-Guti{\'e}rrez}, {Myers}, {Nadathur}, {Nair}, {Nandra}, {do Nascimento}, {Nevin}, {Newman}, {Nidever}, {Nitschelm}, {Noterdaeme}, {O'Connell}, {Olmstead}, {Oravetz}, {Oravetz}, {Osorio}, {Pace}, {Padilla}, {Palanque-Delabrouille}, {Palicio}, {Pan}, {Pan}, {Parker}, {Paviot}, {Peirani}, {Ram{\'r}ez}, {Penny}, {Percival}, {Perez-Fournon}, {P{\'e}rez-R{\`a}fols}, {Petitjean}, {Pieri}, {Pinsonneault}, {Poovelil}, {Povick}, {Prakash}, {Price-Whelan}, {Raddick}, {Raichoor}, {Ray}, {Rembold}, {Rezaie}, {Riffel}, {Riffel}, {Rix}, {Robin}, {Roman-Lopes}, {Rom{\'a}n-Z{\'u}{\~n}iga}, {Rose}, {Ross}, {Rossi}, {Rowlands}, {Rubin}, {Salvato}, {S{\'a}nchez}, {S{\'a}nchez-Menguiano}, {S{\'a}nchez-Gallego}, {Sayres}, {Schaefer}, {Schiavon}, {Schimoia}, {Schlafly}, {Schlegel}, {Schneider}, {Schultheis}, {Schwope}, {Seo}, {Serenelli}, {Shafieloo}, {Shamsi}, {Shao}, {Shen}, {Shetrone}, {Shirley}, {Aguirre}, {Simon}, {Skrutskie}, {Slosar}, {Smethurst}, {Sobeck}, {Sodi},
  {Souto}, {Stark}, {Stassun}, {Steinmetz}, {Stello}, {Stermer}, {Storchi-Bergmann}, {Streblyanska}, {Stringfellow}, {Stutz}, {Su{\'a}rez}, {Sun}, {Taghizadeh-Popp}, {Talbot}, {Tayar}, {Thakar}, {Theriault}, {Thomas}, {Thomas}, {Tinker}, {Tojeiro}, {Toledo}, {Tremonti}, {Troup}, {Tuttle}, {Unda-Sanzana}, {Valentini}, {Vargas-Gonz{\'a}lez}, {Vargas-Maga{\~n}a}, {V{\'a}zquez-Mata}, {Vivek}, {Wake}, {Wang}, {Weaver}, {Weijmans}, {Wild}, {Wilson}, {Wilson}, {Wolthuis}, {Wood-Vasey}, {Yan}, {Yang}, {Y{\`e}che}, {Zamora}, {Zarrouk}, {Zasowski}, {Zhang}, {Zhao}, {Zhao}, {Zheng}, {Zheng}, {Zhu}, \& {Zou}}]{DR16}
{Ahumada}, R., {Prieto}, C.~A., {Almeida}, A., {et~al.} 2020, \apjs, 249, 3

\bibitem[{{Aihara} {et~al.}(2018){Aihara}, {Arimoto}, {Armstrong}, {Arnouts}, {Bahcall}, {Bickerton}, {Bosch}, {Bundy}, {Capak}, {Chan}, {Chiba}, {Coupon}, {Egami}, {Enoki}, {Finet}, {Fujimori}, {Fujimoto}, {Furusawa}, {Furusawa}, {Goto}, {Goulding}, {Greco}, {Greene}, {Gunn}, {Hamana}, {Harikane}, {Hashimoto}, {Hattori}, {Hayashi}, {Hayashi}, {He{\l}miniak}, {Higuchi}, {Hikage}, {Ho}, {Hsieh}, {Huang}, {Huang}, {Ikeda}, {Imanishi}, {Inoue}, {Iwasawa}, {Iwata}, {Jaelani}, {Jian}, {Kamata}, {Karoji}, {Kashikawa}, {Katayama}, {Kawanomoto}, {Kayo}, {Koda}, {Koike}, {Kojima}, {Komiyama}, {Konno}, {Koshida}, {Koyama}, {Kusakabe}, {Leauthaud}, {Lee}, {Lin}, {Lin}, {Lupton}, {Mandelbaum}, {Matsuoka}, {Medezinski}, {Mineo}, {Miyama}, {Miyatake}, {Miyazaki}, {Momose}, {More}, {More}, {Moritani}, {Moriya}, {Morokuma}, {Mukae}, {Murata}, {Murayama}, {Nagao}, {Nakata}, {Niida}, {Niikura}, {Nishizawa}, {Obuchi}, {Oguri}, {Oishi}, {Okabe}, {Okamoto}, {Okura}, {Ono}, {Onodera}, {Onoue}, {Osato}, {Ouchi}, {Price}, {Pyo},
  {Sako}, {Sawicki}, {Shibuya}, {Shimasaku}, {Shimono}, {Shirasaki}, {Silverman}, {Simet}, {Speagle}, {Spergel}, {Strauss}, {Sugahara}, {Sugiyama}, {Suto}, {Suyu}, {Suzuki}, {Tait}, {Takada}, {Takata}, {Tamura}, {Tanaka}, {Tanaka}, {Tanaka}, {Tanaka}, {Terai}, {Terashima}, {Toba}, {Tominaga}, {Toshikawa}, {Turner}, {Uchida}, {Uchiyama}, {Umetsu}, {Uraguchi}, {Urata}, {Usuda}, {Utsumi}, {Wang}, {Wang}, {Wong}, {Yabe}, {Yamada}, {Yamanoi}, {Yasuda}, {Yeh}, {Yonehara}, \& {Yuma}}]{Aihara18}
{Aihara}, H., {Arimoto}, N., {Armstrong}, R., {et~al.} 2018, \pasj, 70, S4

\bibitem[{{Anand} {et~al.}(2022){Anand}, {Kauffmann}, \& {Nelson}}]{Anand22}
{Anand}, A., {Kauffmann}, G., \& {Nelson}, D. 2022, arXiv e-prints, arXiv:2201.07811

\bibitem[{{Anand} {et~al.}(2021){Anand}, {Nelson}, \& {Kauffmann}}]{Anand21}
{Anand}, A., {Nelson}, D., \& {Kauffmann}, G. 2021, \mnras, 504, 65

\bibitem[{{Astropy Collaboration} {et~al.}(2018){Astropy Collaboration}, {Price-Whelan}, {Sip{\H{o}}cz}, {G{\"u}nther}, {Lim}, {Crawford}, {Conseil}, {Shupe}, {Craig}, {Dencheva}, {Ginsburg}, {VanderPlas}, {Bradley}, {P{\'e}rez-Su{\'a}rez}, {de Val-Borro}, {Aldcroft}, {Cruz}, {Robitaille}, {Tollerud}, {Ardelean}, {Babej}, {Bach}, {Bachetti}, {Bakanov}, {Bamford}, {Barentsen}, {Barmby}, {Baumbach}, {Berry}, {Biscani}, {Boquien}, {Bostroem}, {Bouma}, {Brammer}, {Bray}, {Breytenbach}, {Buddelmeijer}, {Burke}, {Calderone}, {Cano Rodr{\'\i}guez}, {Cara}, {Cardoso}, {Cheedella}, {Copin}, {Corrales}, {Crichton}, {D'Avella}, {Deil}, {Depagne}, {Dietrich}, {Donath}, {Droettboom}, {Earl}, {Erben}, {Fabbro}, {Ferreira}, {Finethy}, {Fox}, {Garrison}, {Gibbons}, {Goldstein}, {Gommers}, {Greco}, {Greenfield}, {Groener}, {Grollier}, {Hagen}, {Hirst}, {Homeier}, {Horton}, {Hosseinzadeh}, {Hu}, {Hunkeler}, {Ivezi{\'c}}, {Jain}, {Jenness}, {Kanarek}, {Kendrew}, {Kern}, {Kerzendorf}, {Khvalko}, {King}, {Kirkby}, {Kulkarni},
  {Kumar}, {Lee}, {Lenz}, {Littlefair}, {Ma}, {Macleod}, {Mastropietro}, {McCully}, {Montagnac}, {Morris}, {Mueller}, {Mumford}, {Muna}, {Murphy}, {Nelson}, {Nguyen}, {Ninan}, {N{\"o}the}, {Ogaz}, {Oh}, {Parejko}, {Parley}, {Pascual}, {Patil}, {Patil}, {Plunkett}, {Prochaska}, {Rastogi}, {Reddy Janga}, {Sabater}, {Sakurikar}, {Seifert}, {Sherbert}, {Sherwood-Taylor}, {Shih}, {Sick}, {Silbiger}, {Singanamalla}, {Singer}, {Sladen}, {Sooley}, {Sornarajah}, {Streicher}, {Teuben}, {Thomas}, {Tremblay}, {Turner}, {Terr{\'o}n}, {van Kerkwijk}, {de la Vega}, {Watkins}, {Weaver}, {Whitmore}, {Woillez}, {Zabalza}, \& {Astropy Contributors}}]{astropy18}
{Astropy Collaboration}, {Price-Whelan}, A.~M., {Sip{\H{o}}cz}, B.~M., {et~al.} 2018, \aj, 156, 123

\bibitem[{{Augustin} {et~al.}(2021){Augustin}, {P{\'e}roux}, {Hamanowicz}, {Kulkarni}, {Rahmani}, \& {Zanella}}]{Augustin21}
{Augustin}, R., {P{\'e}roux}, C., {Hamanowicz}, A., {et~al.} 2021, \mnras, 505, 6195

\bibitem[{{Ayromlou} {et~al.}(2019){Ayromlou}, {Nelson}, {Yates}, {Kauffmann}, \& {White}}]{Ayromlou19}
{Ayromlou}, M., {Nelson}, D., {Yates}, R.~M., {Kauffmann}, G., \& {White}, S. D.~M. 2019, \mnras, 487, 4313

\bibitem[{{Bah{\'e}} {et~al.}(2013){Bah{\'e}}, {McCarthy}, {Balogh}, \& {Font}}]{Bahe13}
{Bah{\'e}}, Y.~M., {McCarthy}, I.~G., {Balogh}, M.~L., \& {Font}, A.~S. 2013, \mnras, 430, 3017

\bibitem[{{Bird} {et~al.}(2017){Bird}, {Garnett}, \& {Ho}}]{bird17}
{Bird}, S., {Garnett}, R., \& {Ho}, S. 2017, \mnras, 466, 2111

\bibitem[{{B{\"o}hringer} \& {Werner}(2010)}]{Boehringer2010}
{B{\"o}hringer}, H. \& {Werner}, N. 2010, \aapr, 18, 127

\bibitem[{{Bouch{\'e}} {et~al.}(2006){Bouch{\'e}}, {Murphy}, {P{\'e}roux}, {Csabai}, \& {Wild}}]{bouche}
{Bouch{\'e}}, N., {Murphy}, M.~T., {P{\'e}roux}, C., {Csabai}, I., \& {Wild}, V. 2006, \mnras, 371, 495

\bibitem[{{Bowen} \& {Chelouche}(2011)}]{Bowen}
{Bowen}, D.~V. \& {Chelouche}, D. 2011, \apj, 727, 47

\bibitem[{{Butsky} {et~al.}(2019){Butsky}, {Burchett}, {Nagai}, {Tremmel}, {Quinn}, \& {Werk}}]{Butsky19}
{Butsky}, I.~S., {Burchett}, J.~N., {Nagai}, D., {et~al.} 2019, \mnras, 490, 4292

\bibitem[{{Cantalupo} {et~al.}(2019){Cantalupo}, {Pezzulli}, {Lilly}, {Marino}, {Gallego}, {Schaye}, {Bacon}, {Feltre}, {Kollatschny}, {Nanayakkara}, {Richard}, {Wendt}, {Wisotzki}, \& {Prochaska}}]{Cantalupo19}
{Cantalupo}, S., {Pezzulli}, G., {Lilly}, S.~J., {et~al.} 2019, \mnras, 483, 5188

\bibitem[{{Cen} \& {Ostriker}(1999)}]{CenOstriker99}
{Cen}, R. \& {Ostriker}, J.~P. 1999, \apjl, 519, L109

\bibitem[{{Churchill} {et~al.}(2000){Churchill}, {Mellon}, {Charlton}, {Jannuzi}, {Kirhakos}, {Steidel}, \& {Schneider}}]{Churchill00}
{Churchill}, C.~W., {Mellon}, R.~R., {Charlton}, J.~C., {et~al.} 2000, \apj, 543, 577

\bibitem[{{Clerc} {et~al.}(2014){Clerc}, {Adami}, {Lieu}, {Maughan}, {Pacaud}, {Pierre}, {Sadibekova}, {Smith}, {Valageas}, {Altieri}, {Benoist}, {Maurogordato}, \& {Willis}}]{Clerc2014}
{Clerc}, N., {Adami}, C., {Lieu}, M., {et~al.} 2014, \mnras, 444, 2723

\bibitem[{{Clerc} {et~al.}(2020){Clerc}, {Kirkpatrick}, {Finoguenov}, {Capasso}, {Comparat}, {Damsted}, {Furnell}, {Kukkola}, {Ider Chitham}, {Merloni}, {Salvato}, {Gueguen}, {Dwelly}, {Collins}, {Saro}, {Erfanianfar}, {Schneider}, {Brownstein}, {Mamon}, {Padilla}, {Jullo}, \& {Bizyaev}}]{Clerc2020}
{Clerc}, N., {Kirkpatrick}, C.~C., {Finoguenov}, A., {et~al.} 2020, \mnras, 497, 3976

\bibitem[{{Clerc} {et~al.}(2016){Clerc}, {Merloni}, {Zhang}, {Finoguenov}, {Dwelly}, {Nandra}, {Collins}, {Dawson}, {Kneib}, {Rozo}, {Rykoff}, {Sadibekova}, {Brownstein}, {Lin}, {Ridl}, {Salvato}, {Schwope}, {Steinmetz}, {Seo}, \& {Tinker}}]{Clerc2016}
{Clerc}, N., {Merloni}, A., {Zhang}, Y.~Y., {et~al.} 2016, \mnras, 463, 4490

\bibitem[{{Croston} {et~al.}(2008){Croston}, {Pratt}, {B{\"o}hringer}, {Arnaud}, {Pointecouteau}, {Ponman}, {Sanderson}, {Temple}, {Bower}, \& {Donahue}}]{Croston08}
{Croston}, J.~H., {Pratt}, G.~W., {B{\"o}hringer}, H., {et~al.} 2008, \aap, 487, 431

\bibitem[{{Das} {et~al.}(2021){Das}, {Mathur}, {Gupta}, \& {Krongold}}]{Das21}
{Das}, S., {Mathur}, S., {Gupta}, A., \& {Krongold}, Y. 2021, \apj, 918, 83

\bibitem[{{Dav{\'e}} {et~al.}(2001){Dav{\'e}}, {Cen}, {Ostriker}, {Bryan}, {Hernquist}, {Katz}, {Weinberg}, {Norman}, \& {O'Shea}}]{Dave01}
{Dav{\'e}}, R., {Cen}, R., {Ostriker}, J.~P., {et~al.} 2001, \apj, 552, 473

\bibitem[{{DESI Collaboration} {et~al.}(2016){DESI Collaboration}, {Aghamousa}, {Aguilar}, {Ahlen}, {Alam}, {Allen}, {Allende Prieto}, {Annis}, {Bailey}, {Balland}, {Ballester}, {Baltay}, {Beaufore}, {Bebek}, {Beers}, {Bell}, {Bernal}, {Besuner}, {Beutler}, {Blake}, {Bleuler}, {Blomqvist}, {Blum}, {Bolton}, {Briceno}, {Brooks}, {Brownstein}, {Buckley-Geer}, {Burden}, {Burtin}, {Busca}, {Cahn}, {Cai}, {Cardiel-Sas}, {Carlberg}, {Carton}, {Casas}, {Castander}, {Cervantes-Cota}, {Claybaugh}, {Close}, {Coker}, {Cole}, {Comparat}, {Cooper}, {Cousinou}, {Crocce}, {Cuby}, {Cunningham}, {Davis}, {Dawson}, {de la Macorra}, {De Vicente}, {Delubac}, {Derwent}, {Dey}, {Dhungana}, {Ding}, {Doel}, {Duan}, {Ealet}, {Edelstein}, {Eftekharzadeh}, {Eisenstein}, {Elliott}, {Escoffier}, {Evatt}, {Fagrelius}, {Fan}, {Fanning}, {Farahi}, {Farihi}, {Favole}, {Feng}, {Fernandez}, {Findlay}, {Finkbeiner}, {Fitzpatrick}, {Flaugher}, {Flender}, {Font-Ribera}, {Forero-Romero}, {Fosalba}, {Frenk}, {Fumagalli}, {Gaensicke}, {Gallo},
  {Garcia-Bellido}, {Gaztanaga}, {Pietro Gentile Fusillo}, {Gerard}, {Gershkovich}, {Giannantonio}, {Gillet}, {Gonzalez-de-Rivera}, {Gonzalez-Perez}, {Gott}, {Graur}, {Gutierrez}, {Guy}, {Habib}, {Heetderks}, {Heetderks}, {Heitmann}, {Hellwing}, {Herrera}, {Ho}, {Holland}, {Honscheid}, {Huff}, {Hutchinson}, {Huterer}, {Hwang}, {Illa Laguna}, {Ishikawa}, {Jacobs}, {Jeffrey}, {Jelinsky}, {Jennings}, {Jiang}, {Jimenez}, {Johnson}, {Joyce}, {Jullo}, {Juneau}, {Kama}, {Karcher}, {Karkar}, {Kehoe}, {Kennamer}, {Kent}, {Kilbinger}, {Kim}, {Kirkby}, {Kisner}, {Kitanidis}, {Kneib}, {Koposov}, {Kovacs}, {Koyama}, {Kremin}, {Kron}, {Kronig}, {Kueter-Young}, {Lacey}, {Lafever}, {Lahav}, {Lambert}, {Lampton}, {Landriau}, {Lang}, {Lauer}, {Le Goff}, {Le Guillou}, {Le Van Suu}, {Lee}, {Lee}, {Leitner}, {Lesser}, {Levi}, {L'Huillier}, {Li}, {Liang}, {Lin}, {Linder}, {Loebman}, {Luki{\'c}}, {Ma}, {MacCrann}, {Magneville}, {Makarem}, {Manera}, {Manser}, {Marshall}, {Martini}, {Massey}, {Matheson}, {McCauley}, {McDonald},
  {McGreer}, {Meisner}, {Metcalfe}, {Miller}, {Miquel}, {Moustakas}, {Myers}, {Naik}, {Newman}, {Nichol}, {Nicola}, {Nicolati da Costa}, {Nie}, {Niz}, {Norberg}, {Nord}, {Norman}, {Nugent}, {O'Brien}, {Oh}, {Olsen}, {Padilla}, {Padmanabhan}, {Padmanabhan}, {Palanque-Delabrouille}, {Palmese}, {Pappalardo}, {P{\^a}ris}, {Park}, {Patej}, {Peacock}, {Peiris}, {Peng}, {Percival}, {Perruchot}, {Pieri}, {Pogge}, {Pollack}, {Poppett}, {Prada}, {Prakash}, {Probst}, {Rabinowitz}, {Raichoor}, {Ree}, {Refregier}, {Regal}, {Reid}, {Reil}, {Rezaie}, {Rockosi}, {Roe}, {Ronayette}, {Roodman}, {Ross}, {Ross}, {Rossi}, {Rozo}, {Ruhlmann-Kleider}, {Rykoff}, {Sabiu}, {Samushia}, {Sanchez}, {Sanchez}, {Schlegel}, {Schneider}, {Schubnell}, {Secroun}, {Seljak}, {Seo}, {Serrano}, {Shafieloo}, {Shan}, {Sharples}, {Sholl}, {Shourt}, {Silber}, {Silva}, {Sirk}, {Slosar}, {Smith}, {Smoot}, {Som}, {Song}, {Sprayberry}, {Staten}, {Stefanik}, {Tarle}, {Sien Tie}, {Tinker}, {Tojeiro}, {Valdes}, {Valenzuela}, {Valluri}, {Vargas-Magana},
  {Verde}, {Walker}, {Wang}, {Wang}, {Weaver}, {Weaverdyck}, {Wechsler}, {Weinberg}, {White}, {Yang}, {Yeche}, {Zhang}, {Zhao}, {Zheng}, {Zhou}, {Zhou}, {Zhu}, {Zou}, \& {Zu}}]{DESI16}
{DESI Collaboration}, {Aghamousa}, A., {Aguilar}, J., {et~al.} 2016, arXiv e-prints, arXiv:1611.00036

\bibitem[{{Diemer}(2018)}]{Diemer18}
{Diemer}, B. 2018, \apjs, 239, 35

\bibitem[{{Donahue} \& {Voit}(2022)}]{Donahue22}
{Donahue}, M. \& {Voit}, G.~M. 2022, \physrep, 973, 1

\bibitem[{{Dutta} {et~al.}(2020){Dutta}, {Fumagalli}, {Fossati}, {Lofthouse}, {Prochaska}, {Arrigoni Battaia}, {Bielby}, {Cantalupo}, {Cooke}, {Murphy}, \& {O'Meara}}]{Dutta20}
{Dutta}, R., {Fumagalli}, M., {Fossati}, M., {et~al.} 2020, \mnras, 499, 5022

\bibitem[{{Eke} {et~al.}(2004){Eke}, {Baugh}, {Cole}, {Frenk}, {Norberg}, {Peacock}, {Baldry}, {Bland-Hawthorn}, {Bridges}, {Cannon}, {Colless}, {Collins}, {Couch}, {Dalton}, {de Propris}, {Driver}, {Efstathiou}, {Ellis}, {Glazebrook}, {Jackson}, {Lahav}, {Lewis}, {Lumsden}, {Maddox}, {Madgwick}, {Peterson}, {Sutherland}, \& {Taylor}}]{Eke04}
{Eke}, V.~R., {Baugh}, C.~M., {Cole}, S., {et~al.} 2004, \mnras, 348, 866

\bibitem[{{Ettori}(2003)}]{Ettori}
{Ettori}, S. 2003, \mnras, 344, L13

\bibitem[{{Ferland} {et~al.}(2017){Ferland}, {Chatzikos}, {Guzm{\'a}n}, {Lykins}, {van Hoof}, {Williams}, {Abel}, {Badnell}, {Keenan}, {Porter}, \& {Stancil}}]{cloudy}
{Ferland}, G.~J., {Chatzikos}, M., {Guzm{\'a}n}, F., {et~al.} 2017, \rmxaa, 53, 385

\bibitem[{{Finoguenov} {et~al.}(2020){Finoguenov}, {Rykoff}, {Clerc}, {Costanzi}, {Hagstotz}, {Ider Chitham}, {Kiiveri}, {Kirkpatrick}, {Capasso}, {Comparat}, {Damsted}, {Dupke}, {Erfanianfar}, {Patrick Henry}, {Kaefer}, {Kneib}, {Lindholm}, {Rozo}, {van Waerbeke}, \& {Weller}}]{Finoguenov20}
{Finoguenov}, A., {Rykoff}, E., {Clerc}, N., {et~al.} 2020, \aap, 638, A114

\bibitem[{{Frebel}(2018)}]{Frebel18}
{Frebel}, A. 2018, Annual Review of Nuclear and Particle Science, 68, 237

\bibitem[{{Fresco} {et~al.}(2020){Fresco}, {P{\'e}roux}, {Merloni}, {Hamanowicz}, \& {Szakacs}}]{Fresco20}
{Fresco}, A.~Y., {P{\'e}roux}, C., {Merloni}, A., {Hamanowicz}, A., \& {Szakacs}, R. 2020, \mnras, 499, 5230

\bibitem[{{Fukugita} \& {Peebles}(2004)}]{Fukugita04}
{Fukugita}, M. \& {Peebles}, P.~J.~E. 2004, \apj, 616, 643

\bibitem[{Gad(2021)}]{pygad21}
Gad, A.~F. 2021, PyGAD: An Intuitive Genetic Algorithm Python Library

\bibitem[{{Girardi} {et~al.}(1993){Girardi}, {Biviano}, {Giuricin}, {Mardirossian}, \& {Mezzetti}}]{Girardi93}
{Girardi}, M., {Biviano}, A., {Giuricin}, G., {Mardirossian}, F., \& {Mezzetti}, M. 1993, \apj, 404, 38

\bibitem[{{Gladders} \& {Yee}(2005)}]{Gladders05}
{Gladders}, M.~D. \& {Yee}, H.~K.~C. 2005, \apjs, 157, 1

\bibitem[{{Gonzalez} {et~al.}(2013){Gonzalez}, {Sivanandam}, {Zabludoff}, \& {Zaritsky}}]{Gonzaleznov13}
{Gonzalez}, A.~H., {Sivanandam}, S., {Zabludoff}, A.~I., \& {Zaritsky}, D. 2013, \apj, 778, 14

\bibitem[{{Gonzalez} {et~al.}(2007{\natexlab{a}}){Gonzalez}, {Zaritsky}, \& {Zabludoff}}]{Gonzales07}
{Gonzalez}, A.~H., {Zaritsky}, D., \& {Zabludoff}, A.~I. 2007{\natexlab{a}}, \apj, 666, 147

\bibitem[{{Gonzalez} {et~al.}(2007{\natexlab{b}}){Gonzalez}, {Zaritsky}, \& {Zabludoff}}]{Planelles13}
{Gonzalez}, A.~H., {Zaritsky}, D., \& {Zabludoff}, A.~I. 2007{\natexlab{b}}, \apj, 666, 147

\bibitem[{{Guo} {et~al.}(2018){Guo}, {Shen}, \& {Wang}}]{Guo18}
{Guo}, H., {Shen}, Y., \& {Wang}, S. 2018, {PyQSOFit: Python code to fit the spectrum of quasars}, Astrophysics Source Code Library

\bibitem[{{Hamanowicz} {et~al.}(2020){Hamanowicz}, {P{\'e}roux}, {Zwaan}, {Rahmani}, {Pettini}, {York}, {Klitsch}, {Augustin}, {Krogager}, {Kulkarni}, {Fresco}, {Biggs}, {Milliard}, \& {Vernet}}]{Hamanowicz20}
{Hamanowicz}, A., {P{\'e}roux}, C., {Zwaan}, M.~A., {et~al.} 2020, \mnras, 492, 2347

\bibitem[{{Hummels} {et~al.}(2017){Hummels}, {Smith}, \& {Silvia}}]{Hummels17}
{Hummels}, C.~B., {Smith}, B.~D., \& {Silvia}, D.~W. 2017, \apj, 847, 59

\bibitem[{{Ider Chitham} {et~al.}(2020){Ider Chitham}, {Comparat}, {Finoguenov}, {Clerc}, {Kirkpatrick}, {Damsted}, {Kukkola}, {Capasso}, {Nandra}, {Merloni}, {Bulbul}, {Rykoff}, {Schneider}, \& {Brownstein}}]{IderChitman20}
{Ider Chitham}, J., {Comparat}, J., {Finoguenov}, A., {et~al.} 2020, \mnras, 499, 4768

\bibitem[{{Jones} \& {Forman}(1999)}]{JonesForman99}
{Jones}, C. \& {Forman}, W. 1999, \apj, 511, 65

\bibitem[{{Kirkpatrick} {et~al.}(2021){Kirkpatrick}, {Clerc}, {Finoguenov}, {Damsted}, {Ider Chitham}, {Kukkola}, {Gueguen}, {Furnell}, {Rykoff}, {Comparat}, {Saro}, {Capasso}, {Padilla}, {Erfanianfar}, {Mamon}, {Collins}, {Merloni}, {Brownstein}, \& {Schneider}}]{Kirkpatrick21}
{Kirkpatrick}, C.~C., {Clerc}, N., {Finoguenov}, A., {et~al.} 2021, \mnras, 503, 5763

\bibitem[{{Kirkpatrick} \& {McNamara}(2015)}]{K&M}
{Kirkpatrick}, C.~C. \& {McNamara}, B.~R. 2015, \mnras, 452, 4361

\bibitem[{{Kraljic} {et~al.}(2022){Kraljic}, {Laigle}, {Pichon}, {Peirani}, {Codis}, {Shim}, {Cadiou}, {Pogosyan}, {Arnouts}, {Pieri}, {Ir{\v{s}}i{\v{c}}}, {Morrison}, {O{\~n}orbe}, {P{\'e}rez-R{\`a}fols}, \& {Dalton}}]{Weave22}
{Kraljic}, K., {Laigle}, C., {Pichon}, C., {et~al.} 2022, arXiv e-prints, arXiv:2201.02606

\bibitem[{{Kravtsov} \& {Borgani}(2012)}]{Kravtsov12}
{Kravtsov}, A.~V. \& {Borgani}, S. 2012, \araa, 50, 353

\bibitem[{{Kravtsov} {et~al.}(2005){Kravtsov}, {Nagai}, \& {Vikhlinin}}]{Kravtsov05}
{Kravtsov}, A.~V., {Nagai}, D., \& {Vikhlinin}, A.~A. 2005, \apj, 625, 588

\bibitem[{{Kunz} {et~al.}(2022){Kunz}, {Jones}, \& {Zhuravleva}}]{Kunz22}
{Kunz}, M.~W., {Jones}, T.~W., \& {Zhuravleva}, I. 2022, arXiv e-prints, arXiv:2205.02489

\bibitem[{{Lagan{\'a}} {et~al.}(2013){Lagan{\'a}}, {Martinet}, {Durret}, {Lima Neto}, {Maughan}, \& {Zhang}}]{Lagan13}
{Lagan{\'a}}, T.~F., {Martinet}, N., {Durret}, F., {et~al.} 2013, \aap, 555, A66

\bibitem[{{Lan} \& {Mo}(2018)}]{Lan&Mo18}
{Lan}, T.-W. \& {Mo}, H. 2018, \apj, 866, 36

\bibitem[{{Lanzetta} \& {Bowen}(1990)}]{Lanzetta90}
{Lanzetta}, K.~M. \& {Bowen}, D. 1990, \apj, 357, 321

\bibitem[{{Lee} {et~al.}(2021){Lee}, {Hwang}, \& {Song}}]{Lee}
{Lee}, J.~C., {Hwang}, H.~S., \& {Song}, H. 2021, \mnras, 503, 4309

\bibitem[{{Liu} {et~al.}(2022){Liu}, {Bulbul}, {Ghirardini}, {Liu}, {Klein}, {Clerc}, {{\"O}zsoy}, {Ramos-Ceja}, {Pacaud}, {Comparat}, {Okabe}, {Bahar}, {Biffi}, {Brunner}, {Br{\"u}ggen}, {Buchner}, {Ider Chitham}, {Chiu}, {Dolag}, {Gatuzz}, {Gonzalez}, {Hoang}, {Lamer}, {Merloni}, {Nandra}, {Oguri}, {Ota}, {Predehl}, {Reiprich}, {Salvato}, {Schrabback}, {Sanders}, {Seppi}, \& {Thibaud}}]{Liu2022}
{Liu}, A., {Bulbul}, E., {Ghirardini}, V., {et~al.} 2022, \aap, 661, A2

\bibitem[{{Locatelli} {et~al.}(2023){Locatelli}, {Ponti}, {Zheng}, {Merloni}, {Becker}, {Comparat}, {Dennerl}, {Freyberg}, {Sasaki}, \& {Yeung}}]{Locatelli2023}
{Locatelli}, N., {Ponti}, G., {Zheng}, X., {et~al.} 2023, arXiv e-prints, arXiv:2310.10715

\bibitem[{{Lopez} {et~al.}(2008){Lopez}, {Barrientos}, {Lira}, {Padilla}, {Gilbank}, {Gladders}, {Maza}, {Tejos}, {Vidal}, \& {Yee}}]{Lopez}
{Lopez}, S., {Barrientos}, L.~F., {Lira}, P., {et~al.} 2008, \apj, 679, 1144

\bibitem[{{Lyke} {et~al.}(2020){Lyke}, {Higley}, {McLane}, {Schurhammer}, {Myers}, {Ross}, {Dawson}, {Chabanier}, {Martini}, {Busca}, {Mas des Bourboux}, {Salvato}, {Streblyanska}, {Zarrouk}, {Burtin}, {Anderson}, {Bautista}, {Bizyaev}, {Brandt}, {Brinkmann}, {Brownstein}, {Comparat}, {Green}, {de la Macorra}, {Mu{\~n}oz Guti{\'e}rrez}, {Hou}, {Newman}, {Palanque-Delabrouille}, {P{\^a}ris}, {Percival}, {Petitjean}, {Rich}, {Rossi}, {Schneider}, {Smith}, {Vivek}, \& {Weaver}}]{SDSSLyke}
{Lyke}, B.~W., {Higley}, A.~N., {McLane}, J.~N., {et~al.} 2020, \apjs, 250, 8

\bibitem[{{Marinacci} {et~al.}(2018){Marinacci}, {Vogelsberger}, {Pakmor}, {Torrey}, {Springel}, {Hernquist}, {Nelson}, {Weinberger}, {Pillepich}, {Naiman}, \& {Genel}}]{Marinacci18}
{Marinacci}, F., {Vogelsberger}, M., {Pakmor}, R., {et~al.} 2018, \mnras, 480, 5113

\bibitem[{{Mas-Ribas} {et~al.}(2017){Mas-Ribas}, {Miralda-Escud{\'e}}, {P{\'e}rez-R{\`a}fols}, {Arinyo-i-Prats}, {Noterdaeme}, {Petitjean}, {Schneider}, {York}, \& {Ge}}]{Mas-Ribas17}
{Mas-Ribas}, L., {Miralda-Escud{\'e}}, J., {P{\'e}rez-R{\`a}fols}, I., {et~al.} 2017, \apj, 846, 4

\bibitem[{{Mathur} {et~al.}(2021){Mathur}, {Gupta}, {Das}, {Krongold}, \& {Nicastro}}]{Mathur21}
{Mathur}, S., {Gupta}, A., {Das}, S., {Krongold}, Y., \& {Nicastro}, F. 2021, \apj, 908, 69

\bibitem[{{McCourt} {et~al.}(2012){McCourt}, {Sharma}, {Quataert}, \& {Parrish}}]{Mccourt12}
{McCourt}, M., {Sharma}, P., {Quataert}, E., \& {Parrish}, I.~J. 2012, \mnras, 419, 3319

\bibitem[{{Merloni}(2012)}]{Merloni12}
{Merloni}, A. 2012, in Science from the Next Generation Imaging and Spectroscopic Surveys, 43

\bibitem[{{Merloni} {et~al.}(2019){Merloni}, {Alexander}, {Banerji}, {Boller}, {Comparat}, {Dwelly}, {Fotopoulou}, {McMahon}, {Nandra}, {Salvato}, {Croom}, {Finoguenov}, {Krumpe}, {Lamer}, {Rosario}, {Schwope}, {Shanks}, {Steinmetz}, {Wisotzki}, \& {Worseck}}]{Merloni19}
{Merloni}, A., {Alexander}, D.~A., {Banerji}, M., {et~al.} 2019, The Messenger, 175, 42

\bibitem[{{Mernier} {et~al.}(2018){Mernier}, {Biffi}, {Yamaguchi}, {Medvedev}, {Simionescu}, {Ettori}, {Werner}, {Kaastra}, {de Plaa}, \& {Gu}}]{Mernier18}
{Mernier}, F., {Biffi}, V., {Yamaguchi}, H., {et~al.} 2018, \ssr, 214, 129

\bibitem[{{Mishra} \& {Muzahid}(2022)}]{Mishra}
{Mishra}, S. \& {Muzahid}, S. 2022, arXiv e-prints, arXiv:2201.08545

\bibitem[{{Mishra} {et~al.}(2023){Mishra}, {Muzahid}, {Dutta}, {Srianand}, \& {Charlton}}]{Mishra23}
{Mishra}, S., {Muzahid}, S., {Dutta}, S., {Srianand}, R., \& {Charlton}, J. 2023, arXiv e-prints, arXiv:2305.05698

\bibitem[{{Naiman} {et~al.}(2018){Naiman}, {Pillepich}, {Springel}, {Ramirez-Ruiz}, {Torrey}, {Vogelsberger}, {Pakmor}, {Nelson}, {Marinacci}, {Hernquist}, {Weinberger}, \& {Genel}}]{Naiman18}
{Naiman}, J.~P., {Pillepich}, A., {Springel}, V., {et~al.} 2018, \mnras, 477, 1206

\bibitem[{{Nandra} {et~al.}(2013){Nandra}, {Barret}, {Barcons}, {Fabian}, {den Herder}, {Piro}, {Watson}, {Adami}, {Aird}, {Afonso}, {Alexander}, {Argiroffi}, {Amati}, {Arnaud}, {Atteia}, {Audard}, {Badenes}, {Ballet}, {Ballo}, {Bamba}, {Bhardwaj}, {Stefano Battistelli}, {Becker}, {De Becker}, {Behar}, {Bianchi}, {Biffi}, {B{\^\i}rzan}, {Bocchino}, {Bogdanov}, {Boirin}, {Boller}, {Borgani}, {Borm}, {Bouch{\'e}}, {Bourdin}, {Bower}, {Braito}, {Branchini}, {Branduardi-Raymont}, {Bregman}, {Brenneman}, {Brightman}, {Br{\"u}ggen}, {Buchner}, {Bulbul}, {Brusa}, {Bursa}, {Caccianiga}, {Cackett}, {Campana}, {Cappelluti}, {Cappi}, {Carrera}, {Ceballos}, {Christensen}, {Chu}, {Churazov}, {Clerc}, {Corbel}, {Corral}, {Comastri}, {Costantini}, {Croston}, {Dadina}, {D'Ai}, {Decourchelle}, {Della Ceca}, {Dennerl}, {Dolag}, {Done}, {Dovciak}, {Drake}, {Eckert}, {Edge}, {Ettori}, {Ezoe}, {Feigelson}, {Fender}, {Feruglio}, {Finoguenov}, {Fiore}, {Galeazzi}, {Gallagher}, {Gandhi}, {Gaspari}, {Gastaldello}, {Georgakakis},
  {Georgantopoulos}, {Gilfanov}, {Gitti}, {Gladstone}, {Goosmann}, {Gosset}, {Grosso}, {Guedel}, {Guerrero}, {Haberl}, {Hardcastle}, {Heinz}, {Alonso Herrero}, {Herv{\'e}}, {Holmstrom}, {Iwasawa}, {Jonker}, {Kaastra}, {Kara}, {Karas}, {Kastner}, {King}, {Kosenko}, {Koutroumpa}, {Kraft}, {Kreykenbohm}, {Lallement}, {Lanzuisi}, {Lee}, {Lemoine-Goumard}, {Lobban}, {Lodato}, {Lovisari}, {Lotti}, {McCharthy}, {McNamara}, {Maggio}, {Maiolino}, {De Marco}, {de Martino}, {Mateos}, {Matt}, {Maughan}, {Mazzotta}, {Mendez}, {Merloni}, {Micela}, {Miceli}, {Mignani}, {Miller}, {Miniutti}, {Molendi}, {Montez}, {Moretti}, {Motch}, {Naz{\'e}}, {Nevalainen}, {Nicastro}, {Nulsen}, {Ohashi}, {O'Brien}, {Osborne}, {Oskinova}, {Pacaud}, {Paerels}, {Page}, {Papadakis}, {Pareschi}, {Petre}, {Petrucci}, {Piconcelli}, {Pillitteri}, {Pinto}, {de Plaa}, {Pointecouteau}, {Ponman}, {Ponti}, {Porquet}, {Pounds}, {Pratt}, {Predehl}, {Proga}, {Psaltis}, {Rafferty}, {Ramos-Ceja}, {Ranalli}, {Rasia}, {Rau}, {Rauw}, {Rea}, {Read}, {Reeves},
  {Reiprich}, {Renaud}, {Reynolds}, {Risaliti}, {Rodriguez}, {Rodriguez Hidalgo}, {Roncarelli}, {Rosario}, {Rossetti}, {Rozanska}, {Rovilos}, {Salvaterra}, {Salvato}, {Di Salvo}, {Sanders}, {Sanz-Forcada}, {Schawinski}, {Schaye}, {Schwope}, {Sciortino}, {Severgnini}, {Shankar}, {Sijacki}, {Sim}, {Schmid}, {Smith}, {Steiner}, {Stelzer}, {Stewart}, {Strohmayer}, {Str{\"u}der}, {Sun}, {Takei}, {Tatischeff}, {Tiengo}, {Tombesi}, {Trinchieri}, {Tsuru}, {Ud-Doula}, {Ursino}, {Valencic}, {Vanzella}, {Vaughan}, {Vignali}, {Vink}, {Vito}, {Volonteri}, {Wang}, {Webb}, {Willingale}, {Wilms}, {Wise}, {Worrall}, {Young}, {Zampieri}, {In't Zand}, {Zane}, {Zezas}, {Zhang}, \& {Zhuravleva}}]{Nandra13}
{Nandra}, K., {Barret}, D., {Barcons}, X., {et~al.} 2013, arXiv e-prints, arXiv:1306.2307

\bibitem[{{Narayanan}(2007)}]{Narayan07}
{Narayanan}, A. 2007, in American Astronomical Society Meeting Abstracts, Vol. 211, American Astronomical Society Meeting Abstracts, 72.04

\bibitem[{{Nelson} {et~al.}(2023){Nelson}, {Pillepich}, {Ayromlou}, {Lee}, {Lehle}, {Rohr}, \& {Truong}}]{Nelson23}
{Nelson}, D., {Pillepich}, A., {Ayromlou}, M., {et~al.} 2023, arXiv e-prints, arXiv:2311.06338

\bibitem[{{Nelson} {et~al.}(2019){Nelson}, {Pillepich}, {Springel}, {Pakmor}, {Weinberger}, {Genel}, {Torrey}, {Vogelsberger}, {Marinacci}, \& {Hernquist}}]{Nelson19}
{Nelson}, D., {Pillepich}, A., {Springel}, V., {et~al.} 2019, \mnras, 490, 3234

\bibitem[{{Nelson} {et~al.}(2018){Nelson}, {Pillepich}, {Springel}, {Weinberger}, {Hernquist}, {Pakmor}, {Genel}, {Torrey}, {Vogelsberger}, {Kauffmann}, {Marinacci}, \& {Naiman}}]{Nelson18}
{Nelson}, D., {Pillepich}, A., {Springel}, V., {et~al.} 2018, \mnras, 475, 624

\bibitem[{{Nelson} {et~al.}(2020){Nelson}, {Sharma}, {Pillepich}, {Springel}, {Pakmor}, {Weinberger}, {Vogelsberger}, {Marinacci}, \& {Hernquist}}]{Nelson20}
{Nelson}, D., {Sharma}, P., {Pillepich}, A., {et~al.} 2020, \mnras, 498, 2391

\bibitem[{{Nestor} {et~al.}(2005){Nestor}, {Turnshek}, \& {Rao}}]{Nestor05}
{Nestor}, D.~B., {Turnshek}, D.~A., \& {Rao}, S.~M. 2005, \apj, 628, 637

\bibitem[{{Nielsen} {et~al.}(2020){Nielsen}, {Kacprzak}, {Pointon}, {Murphy}, {Churchill}, \& {Dav{\'e}}}]{Nielsen20}
{Nielsen}, N.~M., {Kacprzak}, G.~G., {Pointon}, S.~K., {et~al.} 2020, \apj, 904, 164

\bibitem[{{Noterdaeme} {et~al.}(2012){Noterdaeme}, {Petitjean}, {Carithers}, {Paris}, {Font-Ribera}, {Bailey}, {Aubourg}, {Bizyaev}, {Ebelke}, {Finley}, {Ge}, {Malanushenko}, {Malanushenko}, {Miralda-Escude}, {Myers}, {Oravetz}, {Pan}, {Pieri}, {Ross}, {Schneider}, {Simmons}, \& {York}}]{noterdaeme12}
{Noterdaeme}, P., {Petitjean}, P., {Carithers}, W.~C., {et~al.} 2012, VizieR Online Data Catalog, J/A+A/547/L1

\bibitem[{{Parks} {et~al.}(2018){Parks}, {Prochaska}, {Dong}, \& {Cai}}]{Parks18}
{Parks}, D., {Prochaska}, J.~X., {Dong}, S., \& {Cai}, Z. 2018, \mnras, 476, 1151

\bibitem[{{P{\'e}roux} \& {Howk}(2020)}]{PerouxHowk20}
{P{\'e}roux}, C. \& {Howk}, J.~C. 2020, \araa, 58, 363

\bibitem[{{Peroux} {et~al.}(2023){Peroux}, {Merloni}, {Liske}, {Salvato}, {Augustin}, {Balzer}, {Cioni}, {Comparat}, {Driver}, {Fresco}, {Garzilli}, {Hamanowicz}, {Klitsch}, {Kneib}, {Krogager}, {Nelson}, {Richard}, {Schady}, {Shen}, {Szakacs}, {Weng}, {Yang}, \& {ByCycle Team}}]{Peroux23}
{Peroux}, C., {Merloni}, A., {Liske}, J., {et~al.} 2023, The Messenger, 190, 42

\bibitem[{{Peterson} {et~al.}(2001){Peterson}, {Paerels}, {Kaastra}, {Arnaud}, {Reiprich}, {Fabian}, {Mushotzky}, {Jernigan}, \& {Sakelliou}}]{Peterson01}
{Peterson}, J.~R., {Paerels}, F.~B.~S., {Kaastra}, J.~S., {et~al.} 2001, \aap, 365, L104

\bibitem[{{Pillepich} {et~al.}(2018){Pillepich}, {Nelson}, {Hernquist}, {Springel}, {Pakmor}, {Torrey}, {Weinberger}, {Genel}, {Naiman}, {Marinacci}, \& {Vogelsberger}}]{Pillepich18}
{Pillepich}, A., {Nelson}, D., {Hernquist}, L., {et~al.} 2018, \mnras, 475, 648

\bibitem[{{Pillepich} {et~al.}(2019){Pillepich}, {Nelson}, {Springel}, {Pakmor}, {Torrey}, {Weinberger}, {Vogelsberger}, {Marinacci}, {Genel}, {van der Wel}, \& {Hernquist}}]{Pillepich19}
{Pillepich}, A., {Nelson}, D., {Springel}, V., {et~al.} 2019, \mnras, 490, 3196

\bibitem[{{Planck Collaboration} {et~al.}(2020){Planck Collaboration}, {Aghanim}, {Akrami}, {Ashdown}, {Aumont}, {Baccigalupi}, {Ballardini}, {Banday}, {Barreiro}, {Bartolo}, {Basak}, {Battye}, {Benabed}, {Bernard}, {Bersanelli}, {Bielewicz}, {Bock}, {Bond}, {Borrill}, {Bouchet}, {Boulanger}, {Bucher}, {Burigana}, {Butler}, {Calabrese}, {Cardoso}, {Carron}, {Challinor}, {Chiang}, {Chluba}, {Colombo}, {Combet}, {Contreras}, {Crill}, {Cuttaia}, {de Bernardis}, {de Zotti}, {Delabrouille}, {Delouis}, {Di Valentino}, {Diego}, {Dor{\'e}}, {Douspis}, {Ducout}, {Dupac}, {Dusini}, {Efstathiou}, {Elsner}, {En{\ss}lin}, {Eriksen}, {Fantaye}, {Farhang}, {Fergusson}, {Fernandez-Cobos}, {Finelli}, {Forastieri}, {Frailis}, {Fraisse}, {Franceschi}, {Frolov}, {Galeotta}, {Galli}, {Ganga}, {G{\'e}nova-Santos}, {Gerbino}, {Ghosh}, {Gonz{\'a}lez-Nuevo}, {G{\'o}rski}, {Gratton}, {Gruppuso}, {Gudmundsson}, {Hamann}, {Handley}, {Hansen}, {Herranz}, {Hildebrandt}, {Hivon}, {Huang}, {Jaffe}, {Jones}, {Karakci}, {Keih{\"a}nen},
  {Keskitalo}, {Kiiveri}, {Kim}, {Kisner}, {Knox}, {Krachmalnicoff}, {Kunz}, {Kurki-Suonio}, {Lagache}, {Lamarre}, {Lasenby}, {Lattanzi}, {Lawrence}, {Le Jeune}, {Lemos}, {Lesgourgues}, {Levrier}, {Lewis}, {Liguori}, {Lilje}, {Lilley}, {Lindholm}, {L{\'o}pez-Caniego}, {Lubin}, {Ma}, {Mac{\'\i}as-P{\'e}rez}, {Maggio}, {Maino}, {Mandolesi}, {Mangilli}, {Marcos-Caballero}, {Maris}, {Martin}, {Martinelli}, {Mart{\'\i}nez-Gonz{\'a}lez}, {Matarrese}, {Mauri}, {McEwen}, {Meinhold}, {Melchiorri}, {Mennella}, {Migliaccio}, {Millea}, {Mitra}, {Miville-Desch{\^e}nes}, {Molinari}, {Montier}, {Morgante}, {Moss}, {Natoli}, {N{\o}rgaard-Nielsen}, {Pagano}, {Paoletti}, {Partridge}, {Patanchon}, {Peiris}, {Perrotta}, {Pettorino}, {Piacentini}, {Polastri}, {Polenta}, {Puget}, {Rachen}, {Reinecke}, {Remazeilles}, {Renzi}, {Rocha}, {Rosset}, {Roudier}, {Rubi{\~n}o-Mart{\'\i}n}, {Ruiz-Granados}, {Salvati}, {Sandri}, {Savelainen}, {Scott}, {Shellard}, {Sirignano}, {Sirri}, {Spencer}, {Sunyaev}, {Suur-Uski}, {Tauber}, {Tavagnacco},
  {Tenti}, {Toffolatti}, {Tomasi}, {Trombetti}, {Valenziano}, {Valiviita}, {Van Tent}, {Vibert}, {Vielva}, {Villa}, {Vittorio}, {Wandelt}, {Wehus}, {White}, {White}, {Zacchei}, \& {Zonca}}]{Planck18}
{Planck Collaboration}, {Aghanim}, N., {Akrami}, Y., {et~al.} 2020, \aap, 641, A6

\bibitem[{{Popesso} {et~al.}(2015){Popesso}, {Biviano}, {Finoguenov}, {Wilman}, {Salvato}, {Magnelli}, {Gruppioni}, {Pozzi}, {Rodighiero}, {Ziparo}, {Berta}, {Elbaz}, {Dickinson}, {Lutz}, {Altieri}, {Aussel}, {Cimatti}, {Fadda}, {Ilbert}, {Le Floch}, {Nordon}, {Poglitsch}, \& {Xu}}]{Popesso15}
{Popesso}, P., {Biviano}, A., {Finoguenov}, A., {et~al.} 2015, \aap, 574, A105

\bibitem[{{Predehl} {et~al.}(2021){Predehl}, {Andritschke}, {Arefiev}, {Babyshkin}, {Batanov}, {Becker}, {B{\"o}hringer}, {Bogomolov}, {Boller}, {Borm}, {Bornemann}, {Br{\"a}uninger}, {Br{\"u}ggen}, {Brunner}, {Brusa}, {Bulbul}, {Buntov}, {Burwitz}, {Burkert}, {Clerc}, {Churazov}, {Coutinho}, {Dauser}, {Dennerl}, {Doroshenko}, {Eder}, {Emberger}, {Eraerds}, {Finoguenov}, {Freyberg}, {Friedrich}, {Friedrich}, {F{\"u}rmetz}, {Georgakakis}, {Gilfanov}, {Granato}, {Grossberger}, {Gueguen}, {Gureev}, {Haberl}, {H{\"a}lker}, {Hartner}, {Hasinger}, {Huber}, {Ji}, {Kienlin}, {Kink}, {Korotkov}, {Kreykenbohm}, {Lamer}, {Lomakin}, {Lapshov}, {Liu}, {Maitra}, {Meidinger}, {Menz}, {Merloni}, {Mernik}, {Mican}, {Mohr}, {M{\"u}ller}, {Nandra}, {Nazarov}, {Pacaud}, {Pavlinsky}, {Perinati}, {Pfeffermann}, {Pietschner}, {Ramos-Ceja}, {Rau}, {Reiffers}, {Reiprich}, {Robrade}, {Salvato}, {Sanders}, {Santangelo}, {Sasaki}, {Scheuerle}, {Schmid}, {Schmitt}, {Schwope}, {Shirshakov}, {Steinmetz}, {Stewart}, {Str{\"u}der},
  {Sunyaev}, {Tenzer}, {Tiedemann}, {Tr{\"u}mper}, {Voron}, {Weber}, {Wilms}, \& {Yaroshenko}}]{Predehl12}
{Predehl}, P., {Andritschke}, R., {Arefiev}, V., {et~al.} 2021, \aap, 647, A1

\bibitem[{{Prochaska} {et~al.}(2017){Prochaska}, {Tejos}, {Crighton}, {jnburchett}, {tiffanyhsyu}, {Tuo-Ji}, {marijana777}, {ktirimba}, {jhennawi}, {Cooke}, {O'Meara}, \& {Werk}}]{Prochaska17}
{Prochaska}, J.~X., {Tejos}, N., {Crighton}, N., {et~al.} 2017, {Linetools/Linetools: Third Minor Release}, Zenodo

\bibitem[{{Ramesh} \& {Nelson}(2023)}]{Ramesh23}
{Ramesh}, R. \& {Nelson}, D. 2023, arXiv e-prints, arXiv:2307.11143

\bibitem[{{Reiprich} {et~al.}(2013){Reiprich}, {Basu}, {Ettori}, {Israel}, {Lovisari}, {Molendi}, {Pointecouteau}, \& {Roncarelli}}]{Reiprich13}
{Reiprich}, T.~H., {Basu}, K., {Ettori}, S., {et~al.} 2013, \ssr, 177, 195

\bibitem[{{Revaz} {et~al.}(2008){Revaz}, {Combes}, \& {Salom{\'e}}}]{Revaz08}
{Revaz}, Y., {Combes}, F., \& {Salom{\'e}}, P. 2008, \aap, 477, L33

\bibitem[{{Rosati} {et~al.}(2002){Rosati}, {Borgani}, \& {Norman}}]{Rosati2002}
{Rosati}, P., {Borgani}, S., \& {Norman}, C. 2002, \araa, 40, 539

\bibitem[{Rykoff {et~al.}(2016)Rykoff, Rozo, Hollowood, Bermeo-Hernandez, Jeltema, Mayers, Romer, Rooney, Saro, Cervantes, Wechsler, Wilcox, Abbott, Abdalla, Allam, Annis, Benoit-L{\'{e} }vy, Bernstein, Bertin, Brooks, Burke, Capozzi, Rosell, Kind, Castander, Childress, Collins, Cunha, D'Andrea, da~Costa, Davis, Desai, Diehl, Dietrich, Doel, Evrard, Finley, Flaugher, Fosalba, Frieman, Glazebrook, Goldstein, Gruen, Gruendl, Gutierrez, Hilton, Honscheid, Hoyle, James, Kay, Kuehn, Kuropatkin, Lahav, Lewis, Lidman, Lima, Maia, Mann, Marshall, Martini, Melchior, Miller, Miquel, Mohr, Nichol, Nord, Ogando, Plazas, Reil, Sahl{\'{e}}n, Sanchez, Santiago, Scarpine, Schubnell, Sevilla-Noarbe, Smith, Soares-Santos, Sobreira, Stott, Suchyta, Swanson, Tarle, Thomas, Tucker, Uddin, Viana, Vikram, Walker, \& and}]{redmapper}
Rykoff, E.~S., Rozo, E., Hollowood, D., {et~al.} 2016, The Astrophysical Journal Supplement Series, 224, 1

\bibitem[{{Saeedzadeh} {et~al.}(2023){Saeedzadeh}, {Jung}, {Rennehan}, {Babul}, {Tremmel}, {Quinn}, {Shao}, {Sharma}, {Mayer}, {O'Sullivan}, \& {Loubser}}]{Saeedzadeh23}
{Saeedzadeh}, V., {Jung}, S.~L., {Rennehan}, D., {et~al.} 2023, \mnras [\eprint[arXiv]{2304.03798}]

\bibitem[{{Sarazin}(1986)}]{Sarazin1986}
{Sarazin}, C.~L. 1986, Reviews of Modern Physics, 58, 1

\bibitem[{{Schaye} {et~al.}(2003){Schaye}, {Aguirre}, {Kim}, {Theuns}, {Rauch}, \& {Sargent}}]{Shaye03}
{Schaye}, J., {Aguirre}, A., {Kim}, T.-S., {et~al.} 2003, \apj, 596, 768

\bibitem[{{Schroetter} {et~al.}(2016){Schroetter}, {Bouch{\'e}}, {Wendt}, {Contini}, {Finley}, {Pell{\'o}}, {Bacon}, {Cantalupo}, {Marino}, {Richard}, {Lilly}, {Schaye}, {Soto}, {Steinmetz}, {Straka}, \& {Wisotzki}}]{Schroetter16}
{Schroetter}, I., {Bouch{\'e}}, N., {Wendt}, M., {et~al.} 2016, \apj, 833, 39

\bibitem[{{Schroetter} {et~al.}(2019){Schroetter}, {Bouch{\'e}}, {Zabl}, {Contini}, {Wendt}, {Schaye}, {Mitchell}, {Muzahid}, {Marino}, {Bacon}, {Lilly}, {Richard}, \& {Wisotzki}}]{Schroetter19}
{Schroetter}, I., {Bouch{\'e}}, N.~F., {Zabl}, J., {et~al.} 2019, \mnras, 490, 4368

\bibitem[{{Seyffert} {et~al.}(2013){Seyffert}, {Cooksey}, {Simcoe}, {O'Meara}, {Kao}, \& {Prochaska}}]{Seyffert13}
{Seyffert}, E.~N., {Cooksey}, K.~L., {Simcoe}, R.~A., {et~al.} 2013, \apj, 779, 161

\bibitem[{{Sharma} {et~al.}(2012){Sharma}, {McCourt}, {Quataert}, \& {Parrish}}]{Sharma12}
{Sharma}, P., {McCourt}, M., {Quataert}, E., \& {Parrish}, I.~J. 2012, \mnras, 420, 3174

\bibitem[{{Simionescu} {et~al.}(2009){Simionescu}, {Werner}, {B{\"o}hringer}, {Kaastra}, {Finoguenov}, {Br{\"u}ggen}, \& {Nulsen}}]{Simionescu09}
{Simionescu}, A., {Werner}, N., {B{\"o}hringer}, H., {et~al.} 2009, \aap, 493, 409

\bibitem[{{Simionescu} {et~al.}(2019){Simionescu}, {ZuHone}, {Zhuravleva}, {Churazov}, {Gaspari}, {Nagai}, {Werner}, {Roediger}, {Canning}, {Eckert}, {Gu}, \& {Paerels}}]{Simionescu19}
{Simionescu}, A., {ZuHone}, J., {Zhuravleva}, I., {et~al.} 2019, \ssr, 215, 24

\bibitem[{{Smee} {et~al.}(2013){Smee}, {Gunn}, {Uomoto}, {Roe}, {Schlegel}, {Rockosi}, {Carr}, {Leger}, {Dawson}, {Olmstead}, {Brinkmann}, {Owen}, {Barkhouser}, {Honscheid}, {Harding}, {Long}, {Lupton}, {Loomis}, {Anderson}, {Annis}, {Bernardi}, {Bhardwaj}, {Bizyaev}, {Bolton}, {Brewington}, {Briggs}, {Burles}, {Burns}, {Castander}, {Connolly}, {Davenport}, {Ebelke}, {Epps}, {Feldman}, {Friedman}, {Frieman}, {Heckman}, {Hull}, {Knapp}, {Lawrence}, {Loveday}, {Mannery}, {Malanushenko}, {Malanushenko}, {Merrelli}, {Muna}, {Newman}, {Nichol}, {Oravetz}, {Pan}, {Pope}, {Ricketts}, {Shelden}, {Sandford}, {Siegmund}, {Simmons}, {Smith}, {Snedden}, {Schneider}, {SubbaRao}, {Tremonti}, {Waddell}, \& {York}}]{Smee13}
{Smee}, S.~A., {Gunn}, J.~E., {Uomoto}, A., {et~al.} 2013, \aj, 146, 32

\bibitem[{{Springel} {et~al.}(2018){Springel}, {Pakmor}, {Pillepich}, {Weinberger}, {Nelson}, {Hernquist}, {Vogelsberger}, {Genel}, {Torrey}, {Marinacci}, \& {Naiman}}]{Springel18}
{Springel}, V., {Pakmor}, R., {Pillepich}, A., {et~al.} 2018, \mnras, 475, 676

\bibitem[{{Szakacs} {et~al.}(2023){Szakacs}, {P{\'e}roux}, {Nelson}, {Zwaan}, {Gr{\"u}n}, {Weng}, {Fresco}, {Bollo}, \& {Casavecchia}}]{Szakacs23}
{Szakacs}, R., {P{\'e}roux}, C., {Nelson}, D., {et~al.} 2023, \mnras, 526, 3744

\bibitem[{{Szakacs} {et~al.}(2021){Szakacs}, {P{\'e}roux}, {Zwaan}, {Hamanowicz}, {Klitsch}, {Fresco}, {Augustin}, {Biggs}, {Kulkarni}, \& {Rahmani}}]{Roland21}
{Szakacs}, R., {P{\'e}roux}, C., {Zwaan}, M., {et~al.} 2021, \mnras, 505, 4746

\bibitem[{{Theuns} {et~al.}(1998){Theuns}, {Leonard}, {Efstathiou}, {Pearce}, \& {Thomas}}]{Theuns98}
{Theuns}, T., {Leonard}, A., {Efstathiou}, G., {Pearce}, F.~R., \& {Thomas}, P.~A. 1998, \mnras, 301, 478

\bibitem[{{Tremblay} {et~al.}(2016){Tremblay}, {Oonk}, {Combes}, {Salom{\'e}}, {O'Dea}, {Baum}, {Voit}, {Donahue}, {McNamara}, {Davis}, {McDonald}, {Edge}, {Clarke}, {Galv{\'a}n-Madrid}, {Bremer}, {Edwards}, {Fabian}, {Hamer}, {Li}, {Maury}, {Russell}, {Quillen}, {Urry}, {Sanders}, \& {Wise}}]{Tremblay16}
{Tremblay}, G.~R., {Oonk}, J.~B.~R., {Combes}, F., {et~al.} 2016, \nat, 534, 218

\bibitem[{{Tremmel} {et~al.}(2019){Tremmel}, {Quinn}, {Ricarte}, {Babul}, {Chadayammuri}, {Natarajan}, {Nagai}, {Pontzen}, \& {Volonteri}}]{Tremmel19}
{Tremmel}, M., {Quinn}, T.~R., {Ricarte}, A., {et~al.} 2019, \mnras, 483, 3336

\bibitem[{{Tripp} {et~al.}(1998){Tripp}, {Lu}, \& {Savage}}]{Tripp98}
{Tripp}, T.~M., {Lu}, L., \& {Savage}, B.~D. 1998, \apj, 508, 200

\bibitem[{{Tumlinson} {et~al.}(2017){Tumlinson}, {Peeples}, \& {Werk}}]{Tumlinson17}
{Tumlinson}, J., {Peeples}, M.~S., \& {Werk}, J.~K. 2017, \araa, 55, 389

\bibitem[{{van de Voort} {et~al.}(2019){van de Voort}, {Springel}, {Mandelker}, {van den Bosch}, \& {Pakmor}}]{Freeke19}
{van de Voort}, F., {Springel}, V., {Mandelker}, N., {van den Bosch}, F.~C., \& {Pakmor}, R. 2019, \mnras, 482, L85

\bibitem[{{Weinberger} {et~al.}(2018){Weinberger}, {Springel}, {Pakmor}, {Nelson}, {Genel}, {Pillepich}, {Vogelsberger}, {Marinacci}, {Naiman}, {Torrey}, \& {Hernquist}}]{Weinberger18}
{Weinberger}, R., {Springel}, V., {Pakmor}, R., {et~al.} 2018, \mnras, 479, 4056

\bibitem[{{Wen} {et~al.}(2012){Wen}, {Han}, \& {Liu}}]{Wen12}
{Wen}, Z.~L., {Han}, J.~L., \& {Liu}, F.~S. 2012, \apjs, 199, 34

\bibitem[{{XRISM Science Team}(2022)}]{XRISM22}
{XRISM Science Team}. 2022, arXiv e-prints, arXiv:2202.05399

\bibitem[{{Yoon} \& {Putman}(2013)}]{Yoon}
{Yoon}, J.~H. \& {Putman}, M.~E. 2013, \apjl, 772, L29

\bibitem[{{Yun} {et~al.}(2019){Yun}, {Pillepich}, {Zinger}, {Nelson}, {Donnari}, {Joshi}, {Rodriguez-Gomez}, {Genel}, {Weinberger}, {Vogelsberger}, \& {Hernquist}}]{Yun19}
{Yun}, K., {Pillepich}, A., {Zinger}, E., {et~al.} 2019, \mnras, 483, 1042

\bibitem[{{Zhu} {et~al.}(2014){Zhu}, {M{\'e}nard}, {Bizyaev}, {Brewington}, {Ebelke}, {Ho}, {Kinemuchi}, {Malanushenko}, {Malanushenko}, {Marchante}, {More}, {Oravetz}, {Pan}, {Petitjean}, \& {Simmons}}]{Zhu14}
{Zhu}, G., {M{\'e}nard}, B., {Bizyaev}, D., {et~al.} 2014, \mnras, 439, 3139

\bibitem[{{Zhu} {et~al.}(2015){Zhu}, {Comparat}, {Kneib}, {Delubac}, {Raichoor}, {Dawson}, {Newman}, {Y{\`e}che}, {Zhou}, \& {Schneider}}]{Zhu15}
{Zhu}, G.~B., {Comparat}, J., {Kneib}, J.-P., {et~al.} 2015, \apj, 815, 48

\bibitem[{{Zou} {et~al.}(2021){Zou}, {Gao}, {Xu}, {Zhou}, {Ma}, {Zhou}, {Zhang}, {Nie}, {Wang}, \& {Xue}}]{Zou21}
{Zou}, H., {Gao}, J., {Xu}, X., {et~al.} 2021, \apjs, 253, 56

\end{thebibliography}

\end{document}